\documentclass[12pt]{article}
\usepackage{amsmath,amssymb,url,mathrsfs, nicefrac, color}
\usepackage[capposition=bottom]{floatrow}
\usepackage{mwe}

\usepackage[skins]{tcolorbox}
\usepackage{lipsum}
\usepackage{comment}
\definecolor{myred}{RGB}{213,94,0}
\definecolor{mygreen}{RGB}{0,158,115}
\definecolor{myblue}{rgb}{0,0,0.75}
\definecolor{bcblue}{RGB}{0,30,52}

\usepackage{hyperref}
\hypersetup{
  colorlinks   = true,    
	urlcolor     = myblue,    
	  linkcolor    = myblue,    
		citecolor    = myblue, 
		 breaklinks=true,   
           colorlinks=true   
	}
\usepackage{tikz}
\usetikzlibrary{patterns, calc,shapes,decorations.pathreplacing}
\usepackage{pgfplots}
\usepackage{subfig}

\usetikzlibrary{arrows.meta} 
\usepackage{amsthm}
\makeatletter
\def\th@plain{%
\thm@notefont{}
  \itshape 
}
\def\th@definition{%
  \thm@notefont{}
	\normalfont 
}
\makeatother
\usepackage{mathtools}
\usepackage[english]{babel}
\usepackage{color}
\usepackage{bm}
\usepackage{accents}	
\usepackage{graphicx}
\usepackage[semicolon]{natbib}

\theoremstyle{plain}

\newtheorem{proposition}{Proposition}
\newtheorem{corollary}{Corollary}
\newtheorem{claim}{Claim}
\theoremstyle{definition}

\newtheorem*{definition*}{Definition}
\newtheorem*{assumption*}{Assumption}

\newtheorem{conjecture*}{Conjecture}
\newtheorem{example}{Example}

\usepackage{url}

\usepackage[top=1in, bottom=1in, left=1in, right=1in]{geometry}
\usepackage{setspace}
\usepackage[bottom]{footmisc}
\usepackage{enumitem}

\newcommand{\type}{\ensuremath{\theta}}
\newcommand{\typeb}{\ensuremath{t}}

\newcommand{\typemeas}{\ensuremath{F}}

\newcommand{\types}{\ensuremath{\Theta}}
\newcommand{\typesb}{\ensuremath{T}}
\newcommand{\wtp}{\ensuremath{v}}
\newcommand{\wtps}{\ensuremath{V}}
\newcommand{\price}{\ensuremath{p}}
\newcommand{\map}{\ensuremath{\mathcal{\pi}}}

\newcommand{\ps}{\ensuremath{\Pi}}
\newcommand{\cs}{\ensuremath{\text{\textup{U}}}}

\newcommand{\psmin}{\ensuremath{\underline{\Pi}}}

\newcommand{\tsmax}{\ensuremath{\overline{\text{\textup{W}}}}}
\newcommand{\profit}{\ensuremath{\Pi}}
\newcommand{\reals}{\ensuremath{\mathbb{R}}}
\newcommand{\info}{\ensuremath{\mathcal{I}}}
\newcommand{\infop}{\ensuremath{\Psi}}

\newcommand{\high}{\ensuremath{H}} 
\newcommand{\low}{\ensuremath{L}} 
\newcommand{\med}{\ensuremath{M}} 

\newcommand{\typetrpost}{\ensuremath{\tilde{\theta}}}
\newcommand{\belief}{\ensuremath{\mu}}
\newcommand{\prior}{\ensuremath{\belief_0}}
\newcommand{\signal}{\ensuremath{s}}
\newcommand{\signals}{\ensuremath{S}}
\newcommand{\post}{\ensuremath{t}}
\newcommand{\dint}{\ensuremath{\mathrm{d}}}
\newcommand{\eqmset}{\ensuremath{\mathcal{E}}}
\newcommand{\menu}{\ensuremath{\mathcal{M}}}

 \newcommand{\R}{\ensuremath{\mathbb{R}}}
 \newcommand{\E}{\ensuremath{\mathbb{E}}}

\newcommand{\prot}{\ensuremath{\mathcal{P}}}
\newcommand{\stratb}{\ensuremath{\sigma_B}}
\newcommand{\strats}{\ensuremath{\sigma_S}}

\newcommand{\ab}{\ensuremath{a_B}}

\newcommand{\asb}{\ensuremath{A_B}}
\newcommand{\ass}{\ensuremath{A_S}}
\newcommand{\e}{\ensuremath{\varepsilon}}
\pgfplotsset{compat=1.17} 
\begin{document}
\title{Data Provision to an Informed Seller\thanks{Ichihashi: Queen's University, \href{mailto:shotaichihashi@gmail.com}{\texttt{shotaichihashi@gmail.com}}. Smolin: Toulouse School of Economics, University of Toulouse Capitole and CEPR, \href{mailto:alexey.v.smolin@gmail.com}{\texttt{alexey.v.smolin@gmail.com}}.
For valuable suggestions and comments, we would like to thank Nageeb Ali, Ricardo Alonso, James Best, Teck Yong Tan, Jidong Zhou, and especially our discussant Daniele Condorelli, as well as seminar participants at Carnegie Mellon University (Tepper), University of California Irvine, Concordia University, Monash-NTU-RUC joint seminar, University of Illinois Urbana-Champaign, Penn State University, Queen's University,  CETC 2022, EARIE 2022, Conference on Mechanism and Institution Design, 2nd Workshop on Contracts, Incentives and Information in Collegio Carlo Alberto, and TSE Digital Conference 2023. 
Smolin acknowledges funding from the French
National Research Agency (ANR) under the Investments for the Future (Investissements d'Avenir) program (grant
ANR-17-EURE-0010).}}
\author{Shota Ichihashi \and Alex Smolin}
\date{\today}
\maketitle
\thispagestyle{empty}

\begin{abstract}
A monopoly seller is privately and imperfectly informed about the buyer's value of the product. The seller uses information to price discriminate the buyer. A designer offers a mechanism that provides the seller with additional information based on the seller’s report about her type. We establish the impossibility of screening for welfare purposes---i.e., the designer can attain any implementable combination of buyer surplus and seller profit by providing the same signal to all seller types. We use this result to characterize the set of implementable welfare outcomes, study the seller’s incentive to acquire third-party data, and demonstrate the trade-off between buyer surplus and efficiency.

\end{abstract}

\newpage
\section{Introduction}


The use of consumer data has become an important and ubiquitous aspect of the interactions between companies and consumers in the digital economy.
Companies collect personal information and track consumers as they browse entertainment portals, post content on social media, and buy products on e-commerce websites. This data is used to adapt online services directly at the point of collection or later if transferred elsewhere.  
One prominent case is the use of consumer data for price discrimination:
A seller may use data on its customers to learn about their preferences and tailor the prices of services and products via outright personalized pricing, providing discount coupons, or steering customers to a more expensive version of similar products.



As the importance of consumer data grows, various parties---such as policymakers, platforms, and consumers themselves---are attempting to control the flow of consumer data to sellers. 
However, the ability of these parties to control the allocation of data across sellers is likely to be limited because of information asymmetry---i.e., sellers may be privately informed about their market demands and customers. 
For example, a regulator might prefer that sellers within a certain group not use certain data, but the regulator may not know which sellers belong to that group.
This raises questions about the scope of data provision to privately informed sellers and its potential impact on consumer surplus, seller profits, and overall efficiency. 


Motivated by the above discussion, we study a model of data provision to a privately informed seller.
The seller is a monopolist and has a product for sale. 
The buyer has a binary uncertain value for the product, which is either high or low.
The seller's interim belief that the value is high, which we call the seller's type, is her private information.
The seller may improve her pricing by obtaining additional information, which we model as a statistical signal that is informative about the value. 
We take a mechanism-design approach and consider a designer who offers a menu of signals to the seller.
In practice, a menu corresponds to a restriction imposed by a regulator or a platform regarding what data sellers can use.
Given a menu, the seller selects a signal and learns about the buyer's value.
Finally, the seller sets a price and the buyer decides whether to purchase the product.

Our focus is the set of all possible welfare outcomes the designer can implement by offering an arbitrary menu of signals.
If the designer observes the seller's type, the set of possible outcomes becomes the ``surplus triangle," as described by \citet*{bergemann2015limits}---i.e.,
any division of the total surplus between the buyer and the seller can be achieved provided that the total surplus is not higher than the efficient surplus, the buyer 
surplus is nonnegative, and the seller's profit is no lower than the profit she can achieve without additional information.

When the seller's type is private, the designer can no longer attain the surplus triangle.
For example, the first-best buyer surplus, which attains efficiency and gives no extra rents to the seller, requires that different types of sellers obtain different signals.
However, if the designer were to implement such an outcome by offering a menu of those signals, some types would choose signals intended for other types in order to extract surplus from the buyer.

Our first main result shows that the designer cannot effectively screen the seller at all: An outcome is implementable through some menu of signals if and only if it is implementable by providing the same public signal to all seller types. This result illustrates the difficulty of eliciting the seller's private information and providing personalized data. To prove the result, we develop novel structural properties of how personalized data provision can impact prices across different seller types and buyer values. We then use these properties to explicitly construct a public signal that replicates the equilibrium pricing behavior of any menu of signals.

We build on this result and geometrically characterize the set of implementable outcomes as the convex hull of an aggregate surplus function that averages across seller types. We show that not only can the designer focus on public signals, but these signals do not need to be complex: Any implementable outcome can be achieved using a signal with at most three signal realizations, and any extreme implementable outcome requires at most two signal realizations.

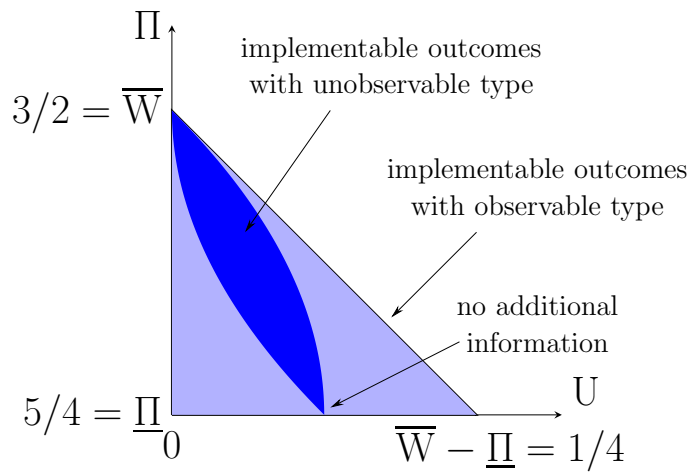
\begin{figure}
\begin{center}

\scalebox{0.9}{
\begin{tikzpicture}
	[xscale=18,yscale=18]
	
\draw[black,-stealth] (0,1.25)--(0.25+0.07,1.25);
 \draw[black,-stealth] (0,1.25)--(0,1.5+0.07);
\node [above right] at (0.25+0.07,1.25) {\Large$\cs$};
\node [left] at (0,1.5+0.07) {\Large$\ps$};
\node [left] at (0,1.5) {\Large$3/2=\tsmax$};
\node [below = 0.1cm] at (0,1.25) {\Large$0$};
\node [below] at (0.25,1.25) {\Large$\hspace{1cm}\tsmax-\psmin=1/4$};
\node [left] at (0,1.25) {\Large$5/4=\psmin$};

\draw[black,thick] 
(0,1.25)--(0,1.5)--(0.25,1.25)--(0,1.25);
\fill[blue!30!white] 
(0,1.25)--(0,1.5)--(0.25,1.25)--(0,1.25);

\node [align=center] at (0.25+0.05,1.5-0.1+0.05) {implementable outcomes};
\node [align=center] at (0.25+0.05,1.47-0.1+0.05) {with observable type};
\draw [-{Stealth[length=2mm, width=1mm]}] (0.2+0.05,1.45-0.1+0.05) -- (0.15+0.03,1.4-0.1+0.03);

\node [align=center] at (0.25+0.05,1.5-0.16) {no additional};
\node [align=center] at (0.25+0.05,1.47-0.16) {information};
\draw [-{Stealth[length=2mm, width=1mm]}] (0.2+0.036,1.45-0.14) -- (0.15-0.02,1.3-0.044);

\fill[blue,thin] 
(0,1.5)--(0.0000490148,1.49505)--(0.000192234,1.4902)--(0.000424168,1.48544)--(0.000739645,1.48077)--(0.00113379,1.47619)--(0.00160199,1.4717)--(0.00213992,1.46729)--(0.00274348,1.46296)--(0.0034088,1.45872)--(0.00413223,1.45455)--(0.00491032,1.45045)--(0.0057398,1.44643)--(0.00661759,1.44248)--(0.00754078,1.4386)--(0.00850662,1.43478)--(0.00951249,1.43103)--(0.0105559,1.42735)--(0.0116346,1.42373)--(0.0127463,1.42017)--(0.0138889,1.41667)--(0.0150604,1.41322)--(0.0162591,1.40984)--(0.017483,1.4065)--(0.0187305,1.40323)--(0.02,1.4)--(0.02129,1.39683)--(0.022599,1.3937)--(0.0239258,1.39063)--(0.0252689,1.3876)--(0.0266272,1.38462)--(0.0279995,1.38168)--(0.0293848,1.37879)--(0.0307818,1.37594)--(0.0321898,1.37313)--(0.0336077,1.37037)--(0.0350346,1.36765)--(0.0364697,1.36496)--(0.0379122,1.36232)--(0.0393613,1.35971)--(0.0408163,1.35714)--(0.0422765,1.35461)--(0.0437413,1.35211)--(0.04521,1.34965)--(0.0466821,1.34722)--(0.048157,1.34483)--(0.0496341,1.34247)--(0.051113,1.34014)--(0.0525931,1.33784)--(0.0540741,1.33557)--(0.0555556,1.33333)--(0.057037,1.33113)--(0.058518,1.32895)--(0.0599983,1.3268)--(0.0614775,1.32468)--(0.0629553,1.32258)--(0.0644313,1.32051)--(0.0659053,1.31847)--(0.067377,1.31646)--(0.0688462,1.31447)--(0.0703125,1.3125)--(0.0717758,1.31056)--(0.0732358,1.30864)--(0.0746923,1.30675)--(0.0761452,1.30488)--(0.0775941,1.30303)--(0.079039,1.3012)--(0.0804798,1.2994)--(0.0819161,1.29762)--(0.0833479,1.29586)--(0.0847751,1.29412)--(0.0861975,1.2924)--(0.0876149,1.2907)--(0.0890274,1.28902)--(0.0904347,1.28736)--(0.0918367,1.28571)--(0.0932335,1.28409)--(0.0946248,1.28249)--(0.0960106,1.2809)--(0.0973908,1.27933)--(0.0987654,1.27778)--(0.100134,1.27624)--(0.101497,1.27473)--(0.102855,1.27322)--(0.104206,1.27174)--(0.105551,1.27027)--(0.106891,1.26882)--(0.108224,1.26738)--(0.109552,1.26596)--(0.110873,1.26455)--(0.112188,1.26316)--(0.113497,1.26178)--(0.1148,1.26042)--(0.116097,1.25907)--(0.117388,1.25773)--(0.118672,1.25641)--(0.11995,1.2551)--(0.121222,1.25381)--(0.122488,1.25253)--(0.123747,1.25126)--(0.125,1.25)--
(0.124997,1.25126)--(0.124987,1.25253)--(0.124971,1.25381)--(0.124948,1.2551)--(0.124918,1.25641)--(0.12488,1.25773)--(0.124836,1.25907)--(0.124783,1.26042)--(0.124722,1.26178)--(0.124654,1.26316)--(0.124577,1.26455)--(0.124491,1.26596)--(0.124396,1.26738)--(0.124292,1.26882)--(0.124178,1.27027)--(0.124055,1.27174)--(0.123921,1.27322)--(0.123777,1.27473)--(0.123623,1.27624)--(0.123457,1.27778)--(0.12328,1.27933)--(0.123091,1.2809)--(0.122889,1.28249)--(0.122676,1.28409)--(0.122449,1.28571)--(0.122209,1.28736)--(0.121955,1.28902)--(0.121687,1.2907)--(0.121405,1.2924)--(0.121107,1.29412)--(0.120794,1.29586)--(0.120465,1.29762)--(0.120119,1.2994)--(0.119756,1.3012)--(0.119376,1.30303)--(0.118977,1.30488)--(0.118559,1.30675)--(0.118122,1.30864)--(0.117665,1.31056)--(0.117188,1.3125)--(0.116688,1.31447)--(0.116167,1.31646)--(0.115623,1.31847)--(0.115056,1.32051)--(0.114464,1.32258)--(0.113847,1.32468)--(0.113204,1.3268)--(0.112535,1.32895)--(0.111837,1.33113)--(0.111111,1.33333)--(0.110355,1.33557)--(0.109569,1.33784)--(0.108751,1.34014)--(0.1079,1.34247)--(0.107015,1.34483)--(0.106096,1.34722)--(0.10514,1.34965)--(0.104146,1.35211)--(0.103114,1.35461)--(0.102041,1.35714)--(0.100926,1.35971)--(0.099769,1.36232)--(0.0985668,1.36496)--(0.0973183,1.36765)--(0.0960219,1.37037)--(0.0946759,1.37313)--(0.0932783,1.37594)--(0.0918274,1.37879)--(0.0903211,1.38168)--(0.0887574,1.38462)--(0.0871342,1.3876)--(0.0854492,1.39063)--(0.0837002,1.3937)--(0.0818846,1.39683)--(0.08,1.4)--(0.0780437,1.40323)--(0.076013,1.4065)--(0.0739049,1.40984)--(0.0717164,1.41322)--(0.0694444,1.41667)--(0.0670857,1.42017)--(0.0646366,1.42373)--(0.0620937,1.42735)--(0.059453,1.43103)--(0.0567108,1.43478)--(0.0538627,1.4386)--(0.0509045,1.44248)--(0.0478316,1.44643)--(0.0446392,1.45045)--(0.0413223,1.45455)--(0.0378756,1.45872)--(0.0342936,1.46296)--(0.0305704,1.46729)--(0.0266999,1.4717)--(0.0226757,1.47619)--(0.0184911,1.48077)--(0.0141389,1.48544)--(0.00961169,1.4902)--(0.00490148,1.49505);

\node [align=center] at (0.25+0.05-0.12,1.5-0.1+0.05+0.1) {implementable outcomes};
\node [align=center] at (0.25+0.05-0.12,1.47-0.1+0.05+0.1) {with unobservable type};
\draw [-{Stealth[length=2mm, width=1mm]}] (0.2+0.05-0.12,1.45-0.1+0.05+0.1) -- (0.15+0.03-0.12,1.4);

\end{tikzpicture}
}
\end{center}
 \caption{The sets of implementable outcomes for observable and unobservable seller types when low value equals 1, high value equals 2, and types are uniformly distributed on $[0,1]$.
 The horizontal axis and vertical axis represent the ex ante expected payoffs of the buyer and the seller, respectively.}\label{fig:intro}
\end{figure}

We use the above results to derive the surplus set in a closed form when the seller's type is uniformly distributed.
\autoref{fig:intro} depicts the set of implementable outcomes for observable types (light blue triangle) and unobservable types (dark blue area).
The seller's private information shrinks the implementable set.
In this example, providing any signal reduces the buyer's ex ante expected payoff because a large mass of seller types will use it to extract surplus from the buyer.
Moreover, the only efficient outcome leads to zero buyer surplus.
The right boundary of the surplus set is spanned by signals that reveal that the buyer has the high value with some probability, and symmetrically, the left boundary is spanned by signals that reveal the low value.


Our result on the impossibility of screening has three implications.
First, the result highlights a tension between consumer protection and efficiency regarding the provision of consumer data to sellers.
The provision of data may enable sellers to tailor pricing, benefit consumers,
and enhance efficiency. 
At the same time, how much and what kind of data sellers should be allowed to use for such beneficial pricing will depend on the specific market conditions and the prior information each seller faces. 
However, in reality, a regulator would not be able to apply different rules to different firms, partly because of the
informational friction we highlight.
In such a situation, allowing some sellers to use a certain piece of information could increase total surplus, but other sellers may use the same information to merely extract consumer surplus.
In some settings, this trade-off can be so stark that it may be impossible to attain efficiency without giving the entire total surplus to the seller.

Second, we study the seller's incentive to acquire third-party data.
Here, third-party data---such as the data a seller may purchase from a data broker---refers to the source of information the seller can acquire but is out of the designer's control.
We can model this as the seller's choice to be privately informed about the buyer.
From the designer's perspective, the seller's private information shrinks the set of implementable outcomes, and in particular, shifts the Pareto frontier downward.
Thus, third-party data typically hurts the designer who cares about social welfare.
In contrast, the seller's incentives to acquire third-party data depend on which implementable outcome is selected.
In particular, if the seller could propose to acquire additional ``first-party" data directly from the buyer but the buyer has the authority to veto the proposal, the seller may prefer not to acquire third-party data. Indeed, because of adverse selection, the buyer may optimally veto acquisition of any additional data by a privately informed seller.
As a result, the third-party data could crowd out the first-party data and reduce the overall information available to the seller.

Third, we show that the surplus set we characterize subsumes the set of equilibrium outcomes of various games in which the buyer and the seller exchange information with each other, including cheap talk communication, voluntary disclosure of the buyer's value, and a request-consent protocol in which a privately informed seller chooses a signal subject to the buyer's consent.
This observation underlies the fact that our mechanism-design approach is useful for evaluating various communication protocols and data collection policies.

Our baseline model assumes that the buyer's value is binary.
On the one hand, this assumption is crucial for our result whereby the designer cannot effectively screen the seller.
On the other hand, many of our key insights extend to the case of general multiple values.
In particular, we show that the seller's private information generally shrinks the set of implementable outcomes and creates a tension between consumer protection and efficiency.
\paragraph*{Related Literature}

First and foremost, our paper relates to the recent theoretical literature on the impact of information under third-degree price discrimination or, equivalently, of market segmentation on market outcomes. In their seminal paper, \cite*{bergemann2015limits} show that all individually rational outcomes can arise in a single-product monopoly setting. Their analysis was later extended to multiproduct markets (\cite*{haghpanah2022limits, haghpanah2019pareto}); competitive markets (\cite*{shi2020welfare, rhodes2022personalized, elliott2021market}); and two-sided markets (\cite*{cosz22}).\footnote{Relatedly, \cite*{roesler2017buyer} and \cite*{roesler2022multi} analyze the informational impact in a second-degree price discrimination setting.}   
We contribute to this literature by highlighting the importance of  the seller's private information, which is absent in those papers.
The presence of such private information is relevant in practice for regulators and platforms, which aim to control the flow of consumer data but are likely to face information asymmetry vis-a-vis sellers.
 We show that the seller's private information limits possible welfare outcomes and introduces the trade-off between consumer welfare and efficiency.

Second, our paper contributes to the literature on consumer privacy and privacy regulation \citep*{acquisti2016economics, choi2019privacy, fainmesser2022digital, argenziano2020information, bergemann2019economics}. 
We highlight the difficulty of tailoring a privacy regulation to unknown market conditions, assess  the impact of access to third-party data on first-party data collection, and compare the performance of several communication protocols. 
To focus on the role of the seller's private information, we abstract away from other economic forces studied in the above papers, such as the use of data for product selection and service improvement as well as information externalities between consumers.


Our mechanism-design framework to study the regulation of a privately informed monopolist follows that of \cite{baron1982regulating}.
We employ this framework in the context of restricting the use of consumer data by a monopolist.
Similar mechanism-design machinery in the context of information provision has recently been used by \cite*{kolotilin2017persuasion}, \cite*{bergemann2018design}; \cite*{smolin2020disclosure}; and \cite*{yang2022selling}. 
All of these allow for a fully flexible way of designing information, following the Bayesian persuasion literature (\cite*{rayo2010optimal}; \cite*{kage11}). 

\section{Model}
There is a seller and a buyer. The seller has a unit good for sale. The buyer's value $\wtp$ for the good is uncertain and either high or low: $\wtp\in \wtps\triangleq \{ \low,\high\}$ with $\high>\low>0$.
The seller is privately informed about the value.
The seller's private information is captured by $\type \in \types \subseteq [0,1]$ and 
represents the seller’s belief that $\wtp = \high$.
The type is distributed  according to measure $\typemeas \in \Delta([0,1])$.\footnote{Given set $X$, we write $\Delta (X)$ for the set of all probability distributions on $X$.}

To improve her pricing, the seller seeks to acquire additional data.
We model data as a statistical signal that can be arbitrarily informative about the value.
Formally, a \emph{signal} $\info = (\signals, \map)$ consists of a set $\signals$ of signal realizations $\signal$ and a family of distributions $\{\map(\cdot|\wtp)\}_{\wtp\in \wtps}$ over $\signals$.
Where it does not cause confusion, we write conditional distribution $\pi(\cdot|v)$ as $\pi(v)$.

We adopt a mechanism-design approach and assume that signals are provided by a designer. At the outset, the designer posts a menu $\menu$ of signals to the seller.
Then the game between the seller and the buyer proceeds as follows.
First, the nature draws the seller's type $\type$ according to prior distribution $\typemeas$ and the buyer's value $v$ according to $\theta$.\footnote{By Bayes' rule, this timing is equivalent to one in which the value is drawn before the type.}
Second, the seller privately observes her type $\type$ and chooses a signal $\info 
 = (S, \pi)\in \menu$.
Third, the seller observes  signal realization $\signal$ drawn according to $\pi(v)$ and posts a price $p\in\R$ for the product.
Finally, the buyer observes value $\wtp$ and price $p$ and decides whether to buy the product.
If the trade occurs, the buyer obtains payoff $v - p$ and the seller obtains payoff $p$.
Otherwise, both players obtain zero payoffs. 

For any given menu $\mathcal{M}$ of signals, the solution concept is a perfect Bayesian equilibrium. 
Any equilibrium induces an \emph{allocation rule} $a:\wtps\rightarrow[0,1]\times\reals$, which specifies for each  value the  probability of a trade and the expected payment from the buyer to the seller. 
We call the corresponding ex ante expected payoffs of the buyer and the seller as the \emph{buyer surplus} and \emph{seller profit}, respectively.
A welfare outcome, or simply \emph{outcome}, refers to a pair of buyer surplus and seller profit.
Each allocation rule leads to a unique outcome, but a given outcome may come from multiple allocation rules. 
An allocation rule and an outcome are \emph{implementable} if they can arise in an equilibrium of some menu.

Our goal is twofold. 
First, we study how the seller's private information limits the designer's ability to implement certain outcomes.
To do so, in Section 3, we characterize the set of implementable outcomes and compare it with the case in which the seller's type is observed.
Second, we aim to derive implications on data policies for regulators and sellers when  the sellers may have private information or face heterogeneous demand conditions.
To this end, in Section 4, we use the characterization result to highlight a general tension between consumer protection and efficiency, and we examine the seller's incentives to acquire external data that is outside the control of the designer. 
Finally, we show that the set of implementable outcomes we characterize subsumes the equilibrium outcomes of various communication protocols motivated by applications.


\subsection{Discussion of Modeling Assumptions}
Before proceeding with the analysis, we briefly discuss several modeling choices that are characteristic of our model.

\paragraph{Data as a Signal.} 
As is common in the literature on information design, we model data as a statistical signal that is informative about the buyer's value. 
This approach bypasses the technical aspects of data analysis and algorithmic implementation and focuses directly on the seller's economic assessments of values.
Different kinds of data, such as web cookies, geolocation data, and consumer behavior on the seller's website, are indicative of consumer preferences and thus can be considered to be signals.
In turn, a signal realization refers to the specific instance of data, such as the contents of a consumer's cookies, his geolocation, or his web-browsing history.
If we view the model as consisting of a continuum of buyers with heterogeneous values and sellers with heterogeneous types, any given signal induces a market segmentation with buyers belonging to the same segment if they have the same realized signal. However, since sellers have different types, the same signal can lead to different market segmentations depending on their types.

\paragraph{Role of the Designer.} 
We abstract away from the designer's objective and focus on all implementable outcomes.
This enables us to make predictions that do not depend on the designer's preferences.
This approach is also useful when considering a player who controls consumer data with a particular objective.
For example, the designer may be a regulator trying to maximize consumer welfare by restricting the data that sellers can use about consumers.
Alternatively, the designer could be a platform or an information intermediary that provides a seller with information about buyers.
Specifying the designer's objective or a specific data collection protocol will select certain implementable outcomes.

\paragraph{Informed Seller.}
The seller's private information about the value of the product to the buyer can come from two sources.
First, the seller may have some information about the buyer that is beyond the designer's control.
Information could come from 
technological constraints like knowledge of the consumer's IP address or from uncontrolled consequences of online interactions such as purchasing decisions. 
Second, the seller may be more informed than the designer about the general quality of their product.  
If the seller has a high-quality product, they believe that the consumer's value is more likely to be high.

\paragraph{Price Discrimination.}
We assume that the seller uses data for third-degree price discrimination. 
In practice, while sellers may be reluctant to display different base prices to different consumers, there are at least two common and indirect ways to price discriminate in the digital economy. 
First, sellers may offer personalized discounts to consumers by setting the base price high and changing the size or frequency of discounts to implement discriminatory pricing. 
Second, sellers may offer personalized recommendations that point consumers toward similar products that vary in price. 
By maintaining a large inventory of such products, sellers can effectively engage in price discrimination.

\section{Surplus Set Characterization}
To study how the seller's private information affects implementable outcomes, we begin our analysis with the benchmark case in which the seller's type is observable to the designer.
We then proceed to the primary scenario of unobservable types and present our main results.
\subsection{Observable Seller Type}
First, suppose that the seller's type is deterministic, i.e., $\types=\{\type_0\}$. 
Denote the maximum feasible total surplus by $\tsmax(\type_0)\triangleq \type_0 \high + (1 - \type_0) \low$ and the seller's profit from   optimal uniform pricing by $\psmin(\type_0)\triangleq \max \{\type_0 \high, \low\}$. Clearly, any feasible welfare outcome $(\cs,\profit)$ must belong to the ``surplus triangle" characterized by constraints $\cs \ge 0$, $\profit \ge \psmin(\type_0)$, and $\cs + \profit \le\tsmax(\type_0)$. \cite{bergemann2015limits} demonstrate  that \emph{any} such welfare outcome can be achieved by some signal. 

\begin{claim}\label{claim:bbm}\emph{(Bergemann, Brooks, Morris (2015))}
If the seller type is commonly known to be $\type_0$, then outcome
$(\cs, \profit)$ is implementable if and only if 
$\cs \ge 0$, $\profit \ge \psmin(\type_0)$, and $\cs + \profit \le \tsmax(\type_0)$. 
\end{claim}

This result immediately generalizes to the case in which the seller's type is drawn according to distribution $F$ but the designer observes the realized type. The aggregate set of implementable outcomes is a Minkowski average of the surplus sets for each realized type. For each type $\theta$, by \autoref{claim:bbm}, the implementable outcomes satisfy three linear constraints, two of which  feature type-dependent terms $\psmin(\type)$ and $\tsmax(\type)$.  Denote their aggregate values averaged across types by 
\begin{align}
    \psmin &\triangleq  \int^1_0 \psmin(\type) \dint F(\type),\\
    \tsmax &\triangleq \int^1_0 \tsmax(\type) \dint F(\type).
\end{align}
From the ex ante perspective, $\psmin$ is the profit the seller can guarantee, $\tsmax$ is the maximum feasible total surplus, and 0 is the welfare level the buyer can guarantee. Hence, the aggregate welfare outcome must belong to a triangle outlined by these constraints. 
At the same time, as in the case of a known type, the converse is also true (the omitted proofs are in Appendix).
\begin{claim}\label{claim:observable}\emph{(Observable type)}
If the seller type is commonly known and distributed according to $F$, then   outcome
$(\cs, \profit)$ is implementable if and only if
$\cs \ge 0$, $\profit \ge \psmin$, and $\cs + \profit \le \tsmax$. 
\end{claim}

The result   implies that any efficient outcome is implementable as long as the seller's profit exceeds the profit under no additional data. 
One notable implementable outcome is the buyer-optimal outcome $(\tsmax - \psmin, \psmin)$, under which the allocation rule is efficient but the seller's profit stays at the level of no additional information. 
At this outcome the  buyer obtains all surplus created by the data. As a result, if the seller's type is observable, there is no inherent trade-off between consumer protection and efficiency.

However, the buyer-optimal outcome generally requires that different seller types obtain different signals. When the seller type is her private information this outcome may not be implementable, and the trade-off between consumer surplus and efficiency reappears. To see this, suppose that the seller's type is either $\type_1\in(0,\low/\high)$ or $\type_2\in(\low/\high,1)$---i.e., the optimal uniform price is $\low$ for type $\type_1$ and $\high$ for type $\type_2$.
At the buyer-optimal outcome, type $\type_1$ must not benefit from the data, because it is already willing to set price $\low$. In contrast, for the outcome to be efficient, type $\type_2$ must be provided with an informative signal that reveals that the buyer's value is $\high$ with a positive probability.\footnote{By Bayes plausibility, some signal realization $s$ will cause type $\type_2$ to believe that value $\high$ is even more likely than $\type_2$.
To cause type $\type_2$ to price efficiently, signal realization $s$ must reveal that the value is $\high$.}
However, in that case   type $\type_1$ could strictly benefit from the signal of $\type_2$, which leads to a contradiction. 

This example shows that the seller's private information restricts the set of implementable outcomes.
In the next section, we analyze the structure of the seller's incentive compatibility constraints and characterize the implementable allocation rules and welfare outcomes.


\subsection{Unobservable Seller Type}
From now on we focus on the case in which the seller's type is unobservable and   characterize the set of implementable outcomes. We show that the seller's private information leads to an extreme adverse selection: Any outcome with heterogeneous data allocation, in which different types self-select into different signals, can be implemented by providing all types with the same public signal.

An important class of menus is a class of \emph{direct mechanisms}. A direct mechanism is a menu of \emph{direct signals} indexed by $\type$ so that $\info(\type)$ sends two signal realizations, $S=\{s_L,s_H\}$, and the likelihood functions $\pi(\type)$ are such that each type $\type$ is willing to choose signal $\info(\type)$ and to set prices $p=L$ and $p=H$ after observing signal realizations $s_L$ and $s_H$ of $\info(\type)$, respectively.
Interpreting signal realization $s_v$  as a recommendation to set price $v \in \{\high, \low\}$,
we can view a direct mechanism as sending a value-dependent price recommendation based on the seller's reported type.
The following result shows that the designer can without loss of generality focus on direct mechanisms (the proof is standard and omitted; it follows the revelation principle argument of \cite{myerson1982optimal} and \cite{bergemann2018design}).
\begin{claim}\label{claim:direct}\emph{(Direct Mechanisms)} An allocation rule is implementable if and only if it is implementable by a direct mechanism.
\end{claim}
Given a direct mechanism, we can parameterize the direct signal $\info(\type)$ for each type $\type$ by probabilities $\alpha(\type)$ and $\beta(\type)$ with which the signal sends realization $s_H$ conditional on values $\high$ and $\low$, respectively.
Without loss of generality, we assume $\beta(\type) \ge \alpha (\type)$.
We can then express signal $\info(\type)$ in matrix form as 
\begin{align}\label{eq:signal_type}
    \begin{array}{c|ccc}
    \info(\type) & s_{L} & s_{H} \\
    \hline v=L & 1-\alpha(\type) & \alpha(\type)\\
    v=H & 1-\beta(\type)  & \beta(\type)
    \end{array}.
\end{align}

For truth-telling to be optimal, each type should prefer her own signal $\info(\type)$ to all other alternatives. The value of a signal depends on the type's response to recommendations, which in turn depends on the posterior beliefs the recommendations induce. Because $\beta(\type) \ge \alpha (\type)$,  the posterior belief rank is the same for all types:  The posterior probability of $\wtp = \high$ is higher after observing $s_H$ than after observing $s_L$. As such, no type would be willing to swap the pricing decisions---i.e., set $p=H$ after $s_L$ and set $p=L$ after $s_H$. Hence, the   relevant incentive constraints are those under which the seller misreports the type and follows the recommendation. Type $\type$'s profit after such a deviation to type $\type'$ is
\begin{align}\label{eq:deviation_profit}
    \profit(\type,\type')&\triangleq(1-\type)(1-\alpha(\type'))L+\type((1-\beta(\type'))L+\beta(\type')H)\\
    &=(1-\alpha(\theta'))L+\theta\left(\alpha(\theta')L+\beta(\theta')(H-L)\right).\notag
\end{align}
Incentive compatibility requires that $ \profit(\type,\type)\geq\profit(\type,\type')$ for all $\type,\type'\in\types$. This property must hold in any direct mechanism and allows us to pin down the structural properties of any implementable allocation rule.
\begin{proposition}\label{prop:structural}\emph{(Allocation Properties)}
In any direct mechanism,  for any $\type_1,\type_2,\type_3\in\types$ such that $\type_1<\type_2<\type_3$, the following hold:
\begin{enumerate}[leftmargin=0.55cm]
\item  \emph{(Monotonicity)} $\alpha(\type_1)\leq\alpha(\type_2)$ and $\beta(\type_1)\leq\beta(\type_2)$; and
\item \emph{(Relative Impact)} $(\beta(\type_3)-\beta(\type_2))(\alpha(\type_2)-\alpha(\type_1))\leq(\beta(\type_2)-\beta(\type_1))(\alpha(\type_3)-\alpha(\type_2))$.
\end{enumerate}
\end{proposition}


The equilibrium properties in \autoref{prop:structural} are direct consequences of the seller's local incentive constraints and do not necessarily hold if the seller's type is observable. Both properties can be interpreted in terms of comparative statics of the seller's behavior with respect to her type. The first property means that the higher the type---i.e., the more likely high-value buyers are in the seller's market---the greater the probability of a higher price for all buyers, irrespective of the seller's additional information. In other words, regardless of the data-provision mechanism, high-value buyers impose negative externalities on low-value buyers because the presence of high-value buyers increases the likelihood that low-value buyers will face higher prices.

The second property evaluates the relative price impact of an increase in the seller's type on low-value and high-value buyers, and is more easily interpreted when viewed as ratio monotonicity---i.e.,  $\frac{\alpha(\type_3)-\alpha(\type_2)}{\beta(\type_3)-\beta(\type_2)}\geq \frac{\alpha(\type_2)-\alpha(\type_1)}{\beta(\type_2)-\beta(\type_1)}$.
Essentially, it states that  the seller type's increasing disproportionately affects low-value buyers at higher types compared with lower types. This property follows from the optimal change in the seller's data strategy. Higher seller types value and primarily choose signals in the menu that help identify low-value buyers, because such signals are more likely to change the types' prior behavior. 
These signals pool low-value buyers with high-value buyers at high-price recommendations, thus exacerbating the price impact of the type's increase on those buyers. 
In contrast, lower seller types value and choose signals that help identify high-value buyers. 
These signals pool low-value buyers with high-value buyers at low-price recommendations, thus mitigating the price impact of the type's increase on those buyers.


The allocation properties of \autoref{prop:structural} outline the constraints imposed by the seller's  private information in direct mechanisms. 
By \autoref{claim:direct}, this result applies to any mechanism, with $\alpha(\type)$ and $\beta(\type)$ being interpreted as the equilibrium pricing probabilities of different types.
In particular, it implies that if the designer provides a single public signal to all types, they optimally respond in such a way that the induced pricing behavior conforms with  \autoref{prop:structural}. It turns out that the opposite is also true: The allocation rule of \emph{any} mechanism can be replicated by providing a single public signal.\footnote{\autoref{prop:public} is reminiscent of the equivalence between the experiments and persuasion mechanisms of \cite{kolotilin2017persuasion}. However, there are important differences between the settings and the results. First, in our setting, the seller's private information is correlated with the value, which affects the seller's assessment and response to data. Second, we  obtain a stronger equivalence result, which applies to allocation rules and not just to the seller's interim utility profile. In our setting, this stronger result is necessary for welfare analysis, because the seller's interim utility profile alone does not determine the buyer's surplus.}
\begin{proposition}\label{prop:public}\emph{(Public Signals)} Any implementable allocation rule can be implemented by a  menu with a single signal in it.
\end{proposition}

\begin{proof}
The proof is constructive. Consider any direct mechanism $(\alpha(\type),\beta(\type))_{\type \in \types}$.  We construct a public signal $\hat\info$ that implements the same outcome.
The signal realization space of $\hat\info$ is $S=[0,1]$. The likelihood function $\pi$ is such that for all $x\in\types$, $\Pr(s\leq x\mid v=L)=\alpha(x)$ and $\Pr(s\leq x\mid v=H)=\beta(x)$. Doing so is possible, because \autoref{prop:structural} ensures that functions $\alpha$ and $\beta$ are   increasing and take values in [0,1].\footnote{If $\alpha(\theta)$  and $\beta(\theta)$ are not right continuous, $\pi$ employs their right-continuous modifications.} 

The defining feature of signal $\hat\info$ is that pooling signal realizations $s$ below and above $\theta$ results in signal $\info(\type)= (\alpha(\type),\beta(\type))$.
That is, $\hat\info$ is Blackwell more informative than either of the signals in the original mechanism. 
However, no type can benefit from the extra informativeness of $\hat\info$. 
Indeed, the second property of \autoref{prop:structural} implies that signal $\hat\info$ satisfies a monotone likelihood ratio property.\footnote{The condition 
$(\alpha(\type_3)-\alpha(\type_2))(\beta(\type_2)-\beta(\type_1))\geq (\alpha(\type_2)-\alpha(\type_1))(\beta(\type_3)-\beta(\type_2))$
is the general definition for CDF $\alpha$ dominating CDF $\beta$ in the monotone likelihood ratio order (see Theorem 1.C.5 of \cite{shaked2007stochastic}).} As such, lower signal realizations induce higher posterior beliefs across all types, and 
 for each type, a best response to $\hat\info$  is characterized by a threshold $\tilde s$ such that the type sets price $p=H$ for all $s<\tilde s$ and price $p=L$ for all $s>\tilde s$. By construction, the choice between different thresholds is equivalent to the choice between different signals in the original mechanism. Hence, by the incentive compatibility of the original mechanism,  type \type\ optimally chooses the threshold $\tilde s=\type$. The resulting allocation rule mimics the allocation rule in the original direct mechanism type by type.
\end{proof}

\subsection{Implementable Outcome Characterization}
\autoref{prop:public} implies that the set of all implementable outcomes coincides with the set of all outcomes that are implementable by public signals. 
We derive the implementable outcomes under public signals in two steps: First, for each signal, we derive the payoffs of the buyer and the seller conditional on each signal realization but unconditional on the seller’s type.
Second, we then aggregate these payoffs across all signal realizations.

Specifically, for any given signal $\info$ and realization $\signal$, define by $\belief\in[0,1]$ the posterior belief that $\signal$ induces in a hypothetical uninformed seller with a prior belief $\prior\triangleq\E[\type]$, so that $\belief\triangleq\Pr(\wtp=H\mid s, \prior)$. We call this belief $\belief$ a \emph{basic} posterior belief. By Bayes' rule, the same signal realization observed by the seller of type $\type$ results in the posterior belief $\post\triangleq\Pr(\wtp=H\mid s,\type)$ equal to\footnote{\cite*{alca16} use this property to analyze Bayesian persuasion with heterogeneous priors.} 
\begin{align}\label{eq:posterior}
    \post(\belief,\type)=\frac{\type\belief(1-\prior)}{\type\belief(1-\prior)+(1-\type)(1-\belief)\prior}.
\end{align}
The posterior belief increases both in $\belief$ and $\type$, equals zero if either of the arguments equals zero, and equals one if either of the arguments equals one. Formula (\ref{eq:posterior}) applies to all types. Therefore, the basic posterior belief $\belief$ fully determines the distribution of the seller's posterior beliefs. The higher the $\belief$, the higher the corresponding distribution of the seller's posterior beliefs in the sense of first-order stochastic dominance---i.e., the beliefs of different seller types move in concordance upon observing the same signal realization.

In turn, the seller's posterior belief determines her pricing decision: The seller sets price $p=H$ if $\post>L/H$ and price $p=L$ if $\post<L/H$. Equivalently, when observing a signal that induces a basic posterior belief $\belief$, the informed seller with type $\type$ sets a high price if $\type>\typetrpost(\belief)$ and a low price if $\type<\typetrpost(\belief)$, where the threshold type $\typetrpost$ is uniquely defined by the condition $\post(\belief,\typetrpost(\belief)))=L/H$. 

By Bayes' consistency, if the seller's posterior belief is $\post$, then the probability that the consumer's value is $\high$ is indeed $\post$. Consequently, the players' expected payoffs after signal realization $s$ that induces a basic posterior belief $\belief$ can be written as
\begin{align}
    \cs(\belief)&=\int_0^{\typetrpost(\belief)}\post(\belief,\type)(H-L)\dint\typemeas(\type),\label{eq:indirect_payoffs_1}\\
    \profit(\belief)&=\int_0^{\typetrpost(\belief)}L\dint\typemeas(\type)+\int_{\typetrpost(\belief)}^{1}\post(\belief,\type)H\dint\typemeas(\type).\label{eq:indirect_payoffs_2}
\end{align}
The welfare functions $\cs(\belief)$ and $\profit(\belief)$ aggregate consumer surplus and seller profit respectively across seller types for any basic posterior belief. We can use these  functions to characterize the set of implementable outcomes. Define by $\mathrm{graph}(\cs,\profit)$ a graph of the vector function that keeps track of players' payoffs $(\cs(\belief),\profit(\belief))$ at different basic beliefs $\belief$. Any public signal is characterized by a distribution of the basic beliefs, which by Bayes' rule must average to the prior belief $\prior$. Therefore, any public signal implements an outcome in a convex hull of $\mathrm{graph}(\cs,\profit)$ with the first component equal to $\prior$. Vice versa, \cite*{auma95} and \cite*{kage11} show that any belief distribution that averages to the prior can be induced by some signal. Therefore, any point in the convex hull of $\mathrm{graph}(\cs,\profit)$ such that the first component is $\prior$ can be implemented by some signal (cf. \cite*{dosm21}).
\begin{proposition}\emph{(Implementable Outcomes)}\label{prop:eqm_outcomes}
The set of all implementable outcomes is
\begin{align}
    \eqmset=\{(x,y):(\prior,x,y)\in\mathrm{co}(\mathrm{graph}(\cs,\profit))\},
\end{align}
where functions $\cs$ and $\profit$ are given by (\ref{eq:indirect_payoffs_1}) and (\ref{eq:indirect_payoffs_2}).
\end{proposition}

For any given type distribution, \autoref{prop:eqm_outcomes} enables a geometric characterization of the set of equilibrium outcomes. 
The result highlights the fact that the type distribution $F$ affects the set of implementable outcomes via its impact on the  indirect payoffs (\ref{eq:indirect_payoffs_1}) and (\ref{eq:indirect_payoffs_2}). 

To simplify exposition, for the rest of this section we assume that $\type$ is continuously distributed over $[0,1]$. The indirect payoffs are then continuous functions of basic belief $\belief$, because $\typetrpost(\belief)$ is continuous and the types are continuously distributed over $[0,1]$. Hence,
by \autoref{prop:eqm_outcomes}, the set $\eqmset$ is compact and convex, and we can characterize its extreme points via supporting Bayesian persuasion problems. Indeed, any extreme point of $\eqmset$  maximizes some linear combination of buyer surplus and seller profit. This maximization problem corresponds to a Bayesian persuasion problem over a public split of basic posterior beliefs to maximize
\begin{align}
	W^{\lambda}(\belief) \triangleq \lambda_{U} \cs (\belief) + \lambda_\profit \profit(\belief)
\end{align}
for some $(\lambda_U, \lambda_\profit) \in \R^2$. We then obtain the following result:
\begin{proposition}\emph{(Extreme Outcomes)}\label{prop:extreme_outcomes}
If the seller type is continuously distributed over $[0,1]$, then the set of extreme implementable outcomes is spanned by solutions to Bayesian persuasion problems parameterized by $\lambda=(\lambda_U, \lambda_\profit)\in\R^2$:
\begin{align}\label{eq:BP}
	\max_{\tau\in \Delta(\Delta(\wtps))}\E_{\mu \sim \tau} [ W^{\lambda}(\belief) ] \quad \textnormal{subject to} \int_{\Delta(\wtps)} \mu \tau(d\mu) = \mu_0.
\end{align}

\end{proposition}

Propositions \ref{prop:eqm_outcomes} and \ref{prop:extreme_outcomes} imply bounds on the necessary complexity of the signals used. When the type is continuously distributed, the indirect payoffs are continuous functions of basic belief $\belief$ and $\mathrm{graph}(\cs,\profit)$ features a single connected component. Thus, by Fenchel-Bunt's theorem, every point in the convex hull of $\mathrm{graph}(\cs,\profit)$ can be generated by a randomization over at most $|V|-1+2=3$ of its points. Such randomization  corresponds to a signal with at most 3 signal realizations. Moreover, any extreme point of $\eqmset$ is implemented by a solution of a Bayesian persuasion problem that can be set to contain at most $|V|=2$ signal realizations.
\begin{corollary}\emph{(Signal Complexity)}\label{cor:signal_complexity}
If the seller type is continuously distributed over $[0,1]$, then any implementable outcome is implementable by a public signal with at most 3 signal realizations. Any extreme point of the set of implementable outcomes is implementable by a public signal with at most 2 signal realizations.
\end{corollary}

\autoref{cor:signal_complexity} shows that even when there are many seller types,  the ex ante consumer and seller payoffs can be obtained if the seller obtains coarse data that features at most 3 labels. 
Furthermore, extreme outcomes, such as a buyer-optimal outcome, can be obtained with even coarser data that features at most 2 labels.

\subsection{Uniform Type Distribution}

So far we have established the general properties of implementable allocations, characterized the set of implementable outcomes, and placed upper bounds on the complexity of necessary signals.
However, we have been silent with respect to a few important questions, such as when providing some data can benefit the buyer or how to achieve a buyer-optimal outcome. 
To tackle these questions and to further illustrate our results, 
we now consider the special case of the seller's type being uniformly distributed over the unit interval. This case captures a natural benchmark in which all seller types are possible and equally likely.

To state the next result, we introduce two classes of signals.
Consider a signal whose likelihood function in tabular form is
\begin{align}\label{eq:signal_type2}
    \begin{array}{c|ccc}
    \info& s_{L} & s_{H} \\
    \hline v=L & 1-\alpha & \alpha\\
    v=H & 1-\beta  & \beta
    \end{array}
\end{align}
for some $\alpha,\beta \in [0,1]$.
A signal is \emph{high-value flagging} if $\alpha =0$, so that realization $s_H$ can arise only if $\wtp = \high$.
A signal is \emph{low-value flagging} if $\beta=1$, so that realization $s_L$ can arise only if $\wtp = \low$.
Note that the fully informative signal ($\alpha =0$ and $\beta = 1$) and the uninformative signal ($\alpha = \beta =0$) belong to these classes.
\begin{proposition}\emph{(Uniform Types)}\label{propositionUniform}
Let the seller type be uniformly distributed on $[0,1]$. The full-information outcome and the no-information outcome are extreme points of $\eqmset$. 
The left and right boundaries of $\eqmset$ connect these two outcomes and are spanned by low-value and high-value flagging signals, respectively. Moreover, the following hold:
	\begin{enumerate}[leftmargin=0.55cm]
		\item If $\frac{L}{H} \ge \frac{1}{2}$, then the buyer-optimal outcome in  $\eqmset$ is generated by the uninformative signal.
		 
		\item If $\frac{L}{H} < \frac{1}{2}$, then 
		the buyer-optimal outcome  in $\eqmset$ is generated by the high-value flagging signal with flagging rate $\beta=\frac{\high - 2\low}{\high - \low}$.
	\end{enumerate}
\end{proposition}

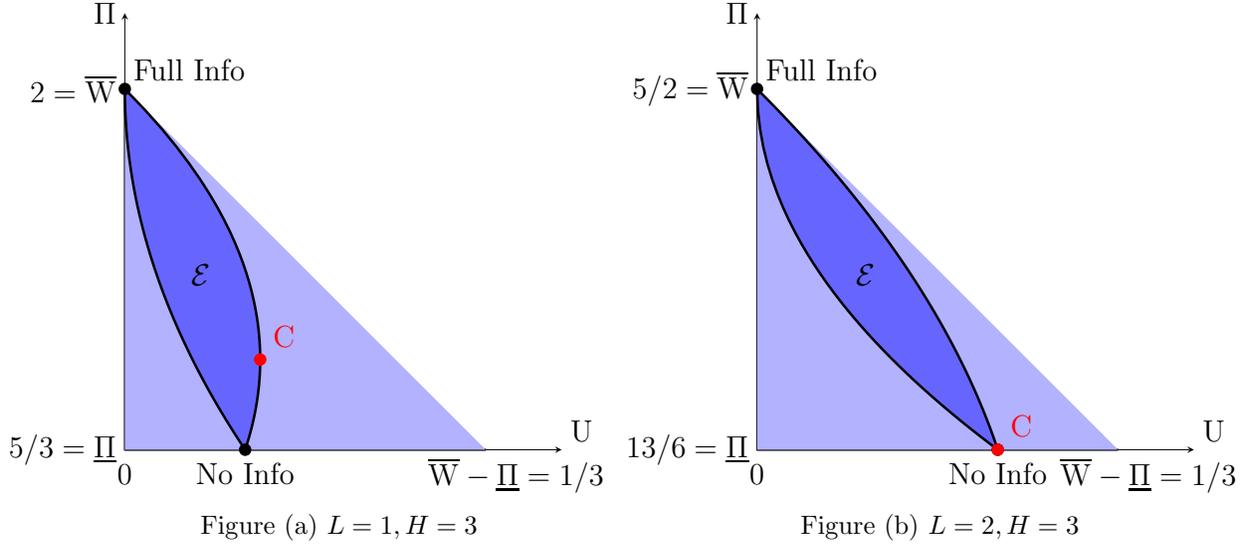
\begin{figure}[t!]
\centering
    \begin{minipage}{0.45\textwidth}
\begin{center}
\subfloat[$L=1, H=3$]{
\scalebox{0.8}{
\hspace{-1.4cm}
\begin{tikzpicture}
	[xscale=18,yscale=18]
\draw[black,-stealth] (0,5/3)--(1/3+0.07,5/3);
 \draw[black,-stealth] (0,5/3)--(0,2+0.07);
\node [above right] at (1/3+0.07,5/3) {\large$\cs$};
\node [left] at (0,2+0.07) {\large$\ps$};
\node [left] at (0,2) {\large$2=\tsmax$};
\node [below = 0.1cm] at (0,5/3) {\large$0$};
\node [below] at (1/3,5/3) {\large$\hspace{1cm}\tsmax-\psmin=1/3$};
\node [left] at (0,5/3) {\large$5/3=\psmin$};

\draw[black,thick] 
(0,2)--(0,5/3)--(1/3,5/3)--(0,5/3);
\fill[blue!30!white] 
(0,5/3)--(0,2)--(1/3,5/3)--(0,5/3);

\fill[blue!60!white,thin] 
(0,2)--(0.0000980296,1.9901)--(0.000384468,1.98039)--(0.000848336,1.97087)--(0.00147929,1.96154)--(0.00226757,1.95238)--(0.00320399,1.9434)--(0.00427985,1.93458)--(0.00548697,1.92593)--(0.00681761,1.91743)--(0.00826446,1.90909)--(0.00982063,1.9009)--(0.0114796,1.89286)--(0.0132352,1.88496)--(0.0150816,1.87719)--(0.0170132,1.86957)--(0.019025,1.86207)--(0.0211118,1.8547)--(0.0232692,1.84746)--(0.0254925,1.84034)--(0.0277778,1.83333)--(0.0301209,1.82645)--(0.0325181,1.81967)--(0.034966,1.81301)--(0.037461,1.80645)--(0.04,1.8)--(0.04258,1.79365)--(0.0451981,1.7874)--(0.0478516,1.78125)--(0.0505378,1.77519)--(0.0532544,1.76923)--(0.0559991,1.76336)--(0.0587695,1.75758)--(0.0615637,1.75188)--(0.0643796,1.74627)--(0.0672154,1.74074)--(0.0700692,1.73529)--(0.0729394,1.72993)--(0.0758244,1.72464)--(0.0787226,1.71942)--(0.0816327,1.71429)--(0.0845531,1.70922)--(0.0874826,1.70423)--(0.0904201,1.6993)--(0.0933642,1.69444)--(0.0963139,1.68966)--(0.0992682,1.68493)--(0.102226,1.68027)--(0.105186,1.67568)--(0.108148,1.67114)--
(0.1111,5/3)--(0.111852,1.66892)--(0.112591,1.67123)--(0.113329,1.67361)--(0.114065,1.67606)--(0.114796,1.67857)--(0.115522,1.68116)--(0.116241,1.68382)--(0.116953,1.68657)--(0.117654,1.68939)--(0.118343,1.69231)--(0.119019,1.69531)--(0.119678,1.69841)--(0.120317,1.70161)--(0.120935,1.70492)--(0.121528,1.70833)--(0.122091,1.71186)--(0.122622,1.71552)--(0.123115,1.7193)--(0.123565,1.72321)--(0.123967,1.72727)--(0.124314,1.73148)--(0.1246,1.73585)--(0.124815,1.74038)--(0.124952,1.7451)--(0.125,1.75)--(0.124948,1.7551)--(0.124783,1.76042)--(0.124491,1.76596)--(0.124055,1.77174)--(0.123457,1.77778)--(0.122676,1.78409)--(0.121687,1.7907)--(0.120465,1.79762)--(0.118977,1.80488)--(0.117188,1.8125)--(0.115056,1.82051)--(0.112535,1.82895)--(0.109569,1.83784)--(0.106096,1.84722)--(0.102041,1.85714)--(0.0973183,1.86765)--(0.0918274,1.87879)--(0.0854492,1.89063)--(0.0780437,1.90323)--(0.0694444,1.91667)--(0.059453,1.93103)--(0.0478316,1.94643)--(0.0342936,1.96296)--(0.0184911,1.98077);

\draw[black,very thick] 
(0,2)--(0.0000980296,1.9901)--(0.000384468,1.98039)--(0.000848336,1.97087)--(0.00147929,1.96154)--(0.00226757,1.95238)--(0.00320399,1.9434)--(0.00427985,1.93458)--(0.00548697,1.92593)--(0.00681761,1.91743)--(0.00826446,1.90909)--(0.00982063,1.9009)--(0.0114796,1.89286)--(0.0132352,1.88496)--(0.0150816,1.87719)--(0.0170132,1.86957)--(0.019025,1.86207)--(0.0211118,1.8547)--(0.0232692,1.84746)--(0.0254925,1.84034)--(0.0277778,1.83333)--(0.0301209,1.82645)--(0.0325181,1.81967)--(0.034966,1.81301)--(0.037461,1.80645)--(0.04,1.8)--(0.04258,1.79365)--(0.0451981,1.7874)--(0.0478516,1.78125)--(0.0505378,1.77519)--(0.0532544,1.76923)--(0.0559991,1.76336)--(0.0587695,1.75758)--(0.0615637,1.75188)--(0.0643796,1.74627)--(0.0672154,1.74074)--(0.0700692,1.73529)--(0.0729394,1.72993)--(0.0758244,1.72464)--(0.0787226,1.71942)--(0.0816327,1.71429)--(0.0845531,1.70922)--(0.0874826,1.70423)--(0.0904201,1.6993)--(0.0933642,1.69444)--(0.0963139,1.68966)--(0.0992682,1.68493)--(0.102226,1.68027)--(0.105186,1.67568)--(0.108148,1.67114)--(0.1111,5/3);

\draw[black,very thick] 
(0.1111,5/3)--(0.111852,1.66892)--(0.112591,1.67123)--(0.113329,1.67361)--(0.114065,1.67606)--(0.114796,1.67857)--(0.115522,1.68116)--(0.116241,1.68382)--(0.116953,1.68657)--(0.117654,1.68939)--(0.118343,1.69231)--(0.119019,1.69531)--(0.119678,1.69841)--(0.120317,1.70161)--(0.120935,1.70492)--(0.121528,1.70833)--(0.122091,1.71186)--(0.122622,1.71552)--(0.123115,1.7193)--(0.123565,1.72321)--(0.123967,1.72727)--(0.124314,1.73148)--(0.1246,1.73585)--(0.124815,1.74038)--(0.124952,1.7451)--(0.125,1.75)--(0.124948,1.7551)--(0.124783,1.76042)--(0.124491,1.76596)--(0.124055,1.77174)--(0.123457,1.77778)--(0.122676,1.78409)--(0.121687,1.7907)--(0.120465,1.79762)--(0.118977,1.80488)--(0.117188,1.8125)--(0.115056,1.82051)--(0.112535,1.82895)--(0.109569,1.83784)--(0.106096,1.84722)--(0.102041,1.85714)--(0.0973183,1.86765)--(0.0918274,1.87879)--(0.0854492,1.89063)--(0.0780437,1.90323)--(0.0694444,1.91667)--(0.059453,1.93103)--(0.0478316,1.94643)--(0.0342936,1.96296)--(0.0184911,1.98077)--(0,2);

\filldraw[black] (0.1111,5/3) circle (0.15pt) node [below = 2pt, align=center] {\large No Info};
\filldraw[black] (0,2) circle (0.15pt) node [above right, align=center] {\large Full Info};

\filldraw[red] (0.125,1.75) circle (0.15pt) node [above right = 2pt, align=center] {\large C};

\draw[black] (0.07,1.85) node [below = 2pt, align=center] {\large \eqmset};
\end{tikzpicture}
    }}
    
    \end{center}
    \end{minipage}
        \begin{minipage}{0.45\textwidth}
\begin{center}
\subfloat[$L=2, H=3$]{
\scalebox{0.8}{
\hspace{-0.6cm}
\begin{tikzpicture}
	[xscale=18,yscale=18]
\draw[black,-stealth] (0,13/6)--(1/3+0.07,13/6);
 \draw[black,-stealth] (0,13/6)--(0,2.5+0.07);
\node [above right] at (1/3+0.07,13/6) {\large$\cs$};
\node [left] at (0,2.5+0.07) {\large$\ps$};
\node [left] at (0,2.5) {\large$5/2=\tsmax$};
\node [below = 0.1cm] at (0,13/6) {\large$0$};
\node [below] at (1/3,13/6) {\large$\hspace{1cm}\tsmax-\psmin=1/3$};
\node [left] at (0,13/6) {\large$13/6=\psmin$};

\draw[black,thick] 
(0,5/2)--(0,13/6)--(1/3,13/6)--(0,13/6);
\fill[blue!30!white] 
(0,13/6)--(0,5/2)--(1/3,13/6)--(0,13/6);

\fill[blue!60!white,thin] 
(0,2.5)--(0.000739645,2.48077)--(0.00274348,2.46296)--(0.0057398,2.44643)--(0.00951249,2.43103)--(0.0138889,2.41667)--(0.0187305,2.40323)--(0.0239258,2.39063)--(0.0293848,2.37879)--(0.0350346,2.36765)--(0.0408163,2.35714)--(0.0466821,2.34722)--(0.0525931,2.33784)--(0.058518,2.32895)--(0.0644313,2.32051)--(0.0703125,2.3125)--(0.0761452,2.30488)--(0.0819161,2.29762)--(0.0876149,2.2907)--(0.0932335,2.28409)--(0.0987654,2.27778)--(0.104206,2.27174)--(0.109552,2.26596)--(0.1148,2.26042)--(0.11995,2.2551)--(0.125,2.25)--(0.12995,2.2451)--(0.1348,2.24038)--(0.139551,2.23585)--(0.144204,2.23148)--(0.14876,2.22727)--(0.153221,2.22321)--(0.157587,2.2193)--(0.161861,2.21552)--(0.166044,2.21186)--(0.170139,2.20833)--(0.174147,2.20492)--(0.17807,2.20161)--(0.18191,2.19841)--(0.185669,2.19531)--(0.189349,2.19231)--(0.192952,2.18939)--(0.19648,2.18657)--(0.199935,2.18382)--(0.203319,2.18116)--(0.206633,2.17857)--(0.209879,2.17606)--(0.213059,2.17361)--(0.216176,2.17123)--(0.219229,2.16892)--
(0.2222,2.1667)--(0.220711,2.17114)--(0.219138,2.17568)--(0.217502,2.18027)--(0.2158,2.18493)--(0.214031,2.18966)--(0.212191,2.19444)--(0.210279,2.1993)--(0.208292,2.20423)--(0.206227,2.20922)--(0.204082,2.21429)--(0.201853,2.21942)--(0.199538,2.22464)--(0.197134,2.22993)--(0.194637,2.23529)--(0.192044,2.24074)--(0.189352,2.24627)--(0.186557,2.25188)--(0.183655,2.25758)--(0.180642,2.26336)--(0.177515,2.26923)--(0.174268,2.27519)--(0.170898,2.28125)--(0.1674,2.2874)--(0.163769,2.29365)--(0.16,2.3)--(0.156087,2.30645)--(0.152026,2.31301)--(0.14781,2.31967)--(0.143433,2.32645)--(0.138889,2.33333)--(0.134171,2.34034)--(0.129273,2.34746)--(0.124187,2.3547)--(0.118906,2.36207)--(0.113422,2.36957)--(0.107725,2.37719)--(0.101809,2.38496)--(0.0956633,2.39286)--(0.0892785,2.4009)--(0.0826446,2.40909)--(0.0757512,2.41743)--(0.0685871,2.42593)--(0.0611407,2.43458)--(0.0533998,2.4434)--(0.0453515,2.45238)--(0.0369822,2.46154)--(0.0282779,2.47087)--(0.0192234,2.48039)--(0.00980296,2.4901);

\draw[black,very thick] 
(0,2.5)--(0.000739645,2.48077)--(0.00274348,2.46296)--(0.0057398,2.44643)--(0.00951249,2.43103)--(0.0138889,2.41667)--(0.0187305,2.40323)--(0.0239258,2.39063)--(0.0293848,2.37879)--(0.0350346,2.36765)--(0.0408163,2.35714)--(0.0466821,2.34722)--(0.0525931,2.33784)--(0.058518,2.32895)--(0.0644313,2.32051)--(0.0703125,2.3125)--(0.0761452,2.30488)--(0.0819161,2.29762)--(0.0876149,2.2907)--(0.0932335,2.28409)--(0.0987654,2.27778)--(0.104206,2.27174)--(0.109552,2.26596)--(0.1148,2.26042)--(0.11995,2.2551)--(0.125,2.25)--(0.12995,2.2451)--(0.1348,2.24038)--(0.139551,2.23585)--(0.144204,2.23148)--(0.14876,2.22727)--(0.153221,2.22321)--(0.157587,2.2193)--(0.161861,2.21552)--(0.166044,2.21186)--(0.170139,2.20833)--(0.174147,2.20492)--(0.17807,2.20161)--(0.18191,2.19841)--(0.185669,2.19531)--(0.189349,2.19231)--(0.192952,2.18939)--(0.19648,2.18657)--(0.199935,2.18382)--(0.203319,2.18116)--(0.206633,2.17857)--(0.209879,2.17606)--(0.213059,2.17361)--(0.216176,2.17123)--(0.219229,2.16892)--(0.2222,2.1667);

\draw[black,very thick] 
(0.2222,2.1667)--(0.220711,2.17114)--(0.219138,2.17568)--(0.217502,2.18027)--(0.2158,2.18493)--(0.214031,2.18966)--(0.212191,2.19444)--(0.210279,2.1993)--(0.208292,2.20423)--(0.206227,2.20922)--(0.204082,2.21429)--(0.201853,2.21942)--(0.199538,2.22464)--(0.197134,2.22993)--(0.194637,2.23529)--(0.192044,2.24074)--(0.189352,2.24627)--(0.186557,2.25188)--(0.183655,2.25758)--(0.180642,2.26336)--(0.177515,2.26923)--(0.174268,2.27519)--(0.170898,2.28125)--(0.1674,2.2874)--(0.163769,2.29365)--(0.16,2.3)--(0.156087,2.30645)--(0.152026,2.31301)--(0.14781,2.31967)--(0.143433,2.32645)--(0.138889,2.33333)--(0.134171,2.34034)--(0.129273,2.34746)--(0.124187,2.3547)--(0.118906,2.36207)--(0.113422,2.36957)--(0.107725,2.37719)--(0.101809,2.38496)--(0.0956633,2.39286)--(0.0892785,2.4009)--(0.0826446,2.40909)--(0.0757512,2.41743)--(0.0685871,2.42593)--(0.0611407,2.43458)--(0.0533998,2.4434)--(0.0453515,2.45238)--(0.0369822,2.46154)--(0.0282779,2.47087)--(0.0192234,2.48039)--(0.00980296,2.4901)--(0,2.5);

\filldraw[black] (0.2222,13/6) circle (0.15pt) node [below = 2pt, align=center] {\large No Info};
\filldraw[red] (0.2222,13/6) circle (0.15pt) node [above right = 2pt, align=center] {\large C};
\filldraw[black] (0,2.5) circle (0.15pt) node [above right, align=center] {\large Full Info};

\draw[black] (0.1 ,2.35) node [below = 2pt, align=center] {\large \eqmset};
\end{tikzpicture}
    }}
   \end{center}
    \end{minipage}
    \caption{Implementable outcomes when seller types are uniformly distributed on $[0,1]$. Light blue denotes the case of observable type.  Dark blue denotes the case of unobservable type. Black boundaries are spanned by flagging signals. Points $C$ indicate buyer-optimal outcomes in $\eqmset$.}\label{fig:uniform}
\end{figure}


The proof of \autoref{propositionUniform} builds on \autoref{prop:extreme_outcomes}. Any extreme point of $\eqmset$ solves a Bayesian persuasion problem \eqref{eq:BP} and is generated by a public signal with two signal realizations. We can thus present the persuasion problem as a   maximization problem with respect to $\alpha$ and $\beta$ and solve it in closed form.

Intuitively, data provision affects the buyer surplus in two ways. On the one hand, it can benefit the buyer by persuading some seller types $\type>L/H$ to set price $p=\low$ even when the buyer has value $\high$.
On the other hand, data provision may harm the buyer if $\type < L/H$, because such seller types would set price $p=\low$ in the absence of data.
Because all implementable outcomes are spanned by public signals, any data provision increases the buyer's surplus for some seller types and decreases it for other types. 
If $L/H$ is high---i.e., a large fraction of seller types set price $\low$ in the absence of data---then the negative effect dominates and any data provision is detrimental for the buyer (Part 1).
If $L/H$ is low, then some data provision is Pareto improving (Part 2).


\autoref{fig:uniform} depicts the sets of implementable outcomes for two concrete cases of value distribution. When the seller type is observable, any individually rational outcome---i.e., the whole surplus triangle---can be implemented by data provision. In contrast, when the seller's type is unobserved, the sets of implementable outcomes shrink.
The uninformative signal implements the  ``south" end of $\eqmset$ as an extreme point, which is also the seller-worst outcome. The right boundary is spanned by high-value flagging and the left boundary by low-value flagging. 
As we move along each boundary from south to north, the corresponding signals have higher flagging rates and become more informative.\footnote{This observation is general---see the proof in Appendix, which characterizes those flagging signals in closed form.}
The two boundaries meet at the ``north" end of $\eqmset$, which is the seller-optimal outcome and implemented by providing full information. 
In fact, this is the only efficient outcome and gives zero surplus to the buyer. 
The buyer-optimal outcome differs in the two cases. 
Figure 2(a) depicts part 2 of the proposition: As $L/H=1/3<1/2$, we have $\beta = 1/2$, so the buyer surplus is maximized by flagging one-half of high-value buyers. 
Figure 2(b) depicts part 1: As $L/H=2/3>1/2$, the buyer surplus is maximized by providing no data. 
In this case, adverse selection is so severe that any additional information would on average hurt the buyer.

\section{Implications}\label{sec:discussion}
We derive three implications of the above results.
First, we show that the seller's private information creates a trade-off between buyer protection and efficiency.
Second, we study the impact of the seller's use of third-party data, which we capture as the seller's choice to acquire private information.
Finally, we demonstrate that the implementable set subsumes the equilibrium outcomes of various games in which the buyer and the seller exchange information with each other.

\subsection{The Trade-off Between Buyer Protection and Efficiency}\label{sec:welfare-efficiency}
If the seller's type is observable, the designer can distribute the efficient total surplus flexibly between the buyer and the seller (\autoref{claim:observable}).
However, implementing an efficient outcome typically requires providing different signals to different types, which is infeasible when the seller's type is her private information (\autoref{prop:public}).
The impossibility of heterogeneous data provision introduces a tension between enhancing consumer surplus and total surplus---and 
in many cases, achieving efficiency implies that the buyer receives no surplus at all.
Below, the seller's ``rent" means the seller's expected profit under a given outcome minus her profit in the absence of additional data.
\begin{proposition}\emph{(Efficiency and Buyer Surplus)}\label{corollaryInefficiency}
Assume that the type distribution is non-degenerate and places probability $1$ on $(0,1)$. 
Then the following hold:
\begin{enumerate}[leftmargin=0.55cm]
\item If there is a positive measure of types strictly above $L/H$, then any efficient outcome gives a strictly positive rent to the seller.
\item If there is a strictly positive measure of types in any neighborhood of $1$, then the only efficient outcome is the outcome in which all seller types obtain full information and perfectly price discriminate the buyer.
\end{enumerate}
\end{proposition}
The intuition is as follows.
To incentivize type $\type> \low/\high$ to price efficiently, a signal has to reveal the buyer's value with some probability $\beta>0$ when $\wtp = \high$, so that in the remaining event, the seller sets price $\low$, believing that the value is likely to be $\low$.
The higher the $\type$, the higher the probability $\beta$ must be and the more informative the signal becomes.
While it is possible for the designer to choose  $\beta$ so that the seller with a known type achieves the same profit as with no additional information, this is not the case for a privately informed seller.
With the privately informed seller, a signal provided to one type for efficient pricing will also be used by lower types to earn positive rents.
Moreover, a type that is arbitrarily close to $1$ must receive the (almost) fully informative signal to price efficiently.
The impossibility of screening implies that all seller types receive full information and the buyer receives no surplus.


The result has an implication for policies regarding the use of consumer data by firms.
Consumer data may enable sellers to tailor pricing, which can benefit consumers and enhance efficiency.
At the same time, how much and what kind of data sellers should be allowed to use for such beneficial pricing will depend on the specific market conditions and the initial endowment of data and technology for each seller. 
However, in practice, a regulator would not be able to tailor a regulation to each firm, partly because of the informational friction highlighted in this paper.
In such a case, the same regulation is applied to heterogeneous sellers.
Our result yields two relevant distortions in such a situation:
First, some sellers obtain too little information, which leads to inefficient pricing.
Second, some sellers obtain too much information, which not only allows them to create surplus but also to extract surplus from consumers.
Our result clarifies this economic force and further shows that the distortion can lead to an extreme outcome in which ensuring efficient pricing erodes the entire buyer surplus.

\subsection{Third-party Data}
In some applications, the seller can obtain third-party data at her will from outside sources.
For example, a seller may learn about buyers from tracking tools or data brokers.
We can analyze the acquisition of such data by interpreting it as the seller's choice to be privately informed.

Specifically, we can study the welfare impact of third-party data by comparing the informed seller---who has private type $\type \sim \typemeas$ with a strictly positive probability of being below and being above $\low/\high$---with the uninformed seller, whose type is deterministic and known to be $\type_0 \triangleq\E_{\tilde\type \sim \typemeas}[\tilde\type]$.
In general, the seller's private information shrinks the set of implementable outcomes (e.g., \autoref{fig:uniform}).
For example, if the type distribution has full support over $[0,1]$, then no efficient outcome, except for the outcome in which the seller extracts full surplus, is implementable when the seller has private information (\autoref{corollaryInefficiency}).

A natural question is how the seller's acquisition of third-party data affects the seller and the buyer.
The answer depends on which implementable outcome is selected.
As an illustration, suppose that 
the selected outcome is the \emph{constrained seller-optimal outcome}---i.e., the implementable outcome that maximizes seller surplus subject to the constraint whereby the buyer must weakly benefit from the data provision in terms of his ex ante expected surplus.
This outcome reflects a commonly observed feature of the digital economy, in which  firms specify the terms of data collection and usage, but harmful data collection may be prevented by a user's opting out or by a regulator's intervention.

\newpage                   
\begin{proposition}\emph{(Third-party Data)}\label{prop:third-party-data}
Let $(\cs, \profit)$ and $(\cs', \profit')$ be the constrained seller-optimal outcomes given the uninformed seller and the informed seller.
Then the following hold:
\begin{enumerate}[leftmargin=0.55cm]
	\item If $\theta_0 > \frac{\low}{\high}$, then $\profit > \profit'$ and $\cs < \cs'$---i.e., the seller is worse off and the buyer is better off when the seller has  private information.
	\item 
If $\theta_0 < \frac{\low}{\high}$, then $\profit < \profit'$ and $\cs > \cs'$---i.e., the seller is better off and the buyer is worse off when the seller has private information.
\end{enumerate}
\end{proposition}

The intuition for Part 1 is as follows.
The condition $\theta_0 > \low/\high$ means that the uninformed seller believes that the buyer's value is likely to be high, so she optimally sets price $\high$, which leads to zero buyer surplus.
In such a case, the constrained seller-optimal outcome is that the seller obtains full information and extracts the efficient total surplus.
In contrast, the constrained seller-optimal outcome for the privately informed seller dictates that the seller never obtains full information, because the buyer surplus at no additional data provision is still positive. 
Consequently, the seller's private information decreases the seller's profit and increases buyer surplus.
Part 1 of the result suggests that the seller may choose not to be privately informed about the buyer in order to eliminate adverse selection and obtain superior first-party data.
This choice, while seemingly privacy friendly, could in fact harm the buyer.\footnote{In other words, third-party data may crowd out first-party data. A similar interplay between several sources of information is the main focus of \cite{alza21} in an investment setting.}

The argument for Part 2 is symmetric:
Given that the uninformed seller sets price $\low$, the constrained seller-optimal outcome is to provide no data.
Compared with this outcome, the seller's private information increases her profit at the expense of buyer surplus and efficiency.

\subsection{Communication Protocols}\label{sec:protocols}
So far, we have adopted a mechanism-design approach and studied the set of implementable outcomes. 
We now show that this set contains the equilibrium outcomes in a large class of communication protocols with an informed buyer and seller, such as cheap-talk communication, voluntary disclosure, and the collection of verifiable data.

Formally, assume that the buyer and seller are  privately informed about value $v$, having types $t$ and $\theta$, respectively, which are independent conditional on $\wtp$. Before trade, the players can communicate via a predefined two-stage protocol.\footnote{This is a large class of  protocols that are natural in applications. However, like a revelation principle, our results can be stated with respect to a broader class of protocols at the expense of additional notation.} 
A \emph{protocol} \prot\ is a triple $(\ass,\{\asb(\typeb)\}_{\typeb\in \typesb},\infop)$. 
The sets $\ass$ and $\{\asb(\typeb)\}_{\typeb\in\typesb}$ denote the seller's action space and the buyer's action space, respectively. The buyer's action space $A_B(t)$ can depend on his type $t$, which enables us to capture, e.g., voluntary disclosure settings.   
The signal scheme is denoted by $\infop=(S,\pi)$.
Here, $S$ is the set of signal realizations the seller can possibly observe in the protocol and  $\pi:\wtps\times\ass\times\asb\rightarrow \Delta(S)$ is the likelihood function of different signal realizations, which can depend on the value and, importantly, on the players' actions. In the protocol, first the seller takes an action. Second, the buyer observes the seller's action and chooses his own action. A seller strategy is  $\strats:\types\rightarrow \Delta(A_S)$ and a buyer   strategy is $\stratb:\typesb\times\ass\rightarrow \Delta(\asb(\typeb))$. At the end of the communication, the seller observes signal realization $s$ according to $\infop$.

After communicating within the protocol, the seller sets the price $p$. The buyer learns the value $\wtp$, observes the price, and decides whether to purchase the good. If trade occurs, the buyer's ex post payoff is $v-p$ and the seller's ex post payoff is $p$. If trade does not occur, both players obtain a zero payoff. The solution concept is a perfect Bayesian equilibrium.

Any equilibrium in a protocol results in  an allocation rule $a:\wtps\rightarrow[0,1]\times\reals$, which as before specifies the  probability of a trade and the expected payment from the buyer to the seller for each  value.
We continue to use (welfare) outcome, buyer surplus, and seller profit to mean relevant ex ante expected payoffs.

\begin{proposition}\label{prop:direct}\emph{(Menu Mechanisms)} An allocation rule can arise in an equilibrium with some communication protocol if and only if it can arise in an equilibrium of a menu mechanism.
\end{proposition}
\begin{proof}
The proof follows the revelation principle argument of \cite{myerson1982optimal,myerson1983mechanism}, but is adapted to fit the specific details of our environment. For the ``if" direction, note that any menu mechanism is itself an example of a communication protocol with $\asb(t)\equiv \{a_0\}$. For the ``only if" direction, consider any protocol and an equilibrium in it. In this equilibrium, each action of the seller induces a signal $\info(a)=(S,\pi(a))$ with $\pi(a):V\rightarrow \Delta(S)$, where the likelihood function averages over the buyer's equilibrium strategy and the signal scheme. We can replace this protocol with a menu mechanism $\mathcal{M}=\{\info(a)\}_{a\in A_S}$. 
This change does not alter the seller's equilibrium strategy and results in the same allocation rule as the original protocol. 
Indeed, the only thing that matters at the trading stage is the seller's estimate of the buyer's value. 
It does not matter whether this estimate is obtained through direct data provision or equilibrium inference. Moreover, the buyer type is not useful for screening the seller type because they are conditionally independent. As a result, menu mechanisms with direct data provision implement all equilibrium allocation rules.
\end{proof}

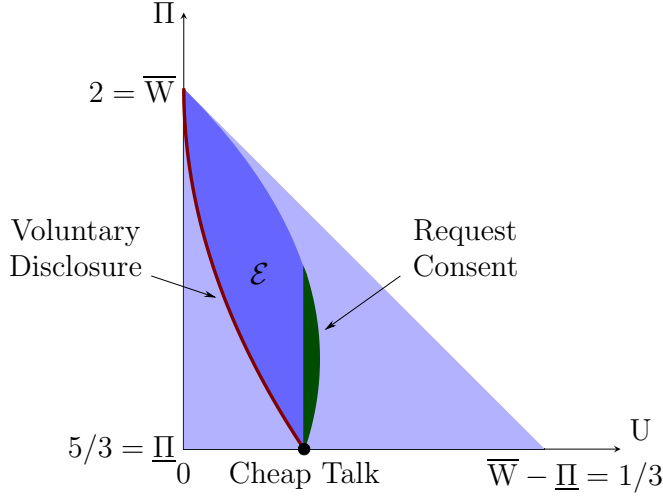
\begin{figure}[t!]
\centering

\scalebox{0.8}{
\hspace{-1.4cm}
\begin{tikzpicture}
	[xscale=18,yscale=18]
\draw[black,-stealth] (0,5/3)--(1/3+0.07,5/3);
 \draw[black,-stealth] (0,5/3)--(0,2+0.07);
\node [above right] at (1/3+0.07,5/3) {\large$\cs$};
\node [left] at (0,2+0.07) {\large$\ps$};
\node [left] at (0,2) {\large$2=\tsmax$};
\node [below = 0.1cm] at (0,5/3) {\large$0$};
\node [below] at (1/3,5/3) {\large$\hspace{1cm}\tsmax-\psmin=1/3$};
\node [left] at (0,5/3) {\large$5/3=\psmin$};

\draw[black,thick] 
(0,2)--(0,5/3)--(1/3,5/3)--(0,5/3);
\fill[blue!30!white] 
(0,5/3)--(0,2)--(1/3,5/3)--(0,5/3);

\fill[blue!60!white] 
(0,2)--(0.0000980296,1.9901)--(0.000384468,1.98039)--(0.000848336,1.97087)--(0.00147929,1.96154)--(0.00226757,1.95238)--(0.00320399,1.9434)--(0.00427985,1.93458)--(0.00548697,1.92593)--(0.00681761,1.91743)--(0.00826446,1.90909)--(0.00982063,1.9009)--(0.0114796,1.89286)--(0.0132352,1.88496)--(0.0150816,1.87719)--(0.0170132,1.86957)--(0.019025,1.86207)--(0.0211118,1.8547)--(0.0232692,1.84746)--(0.0254925,1.84034)--(0.0277778,1.83333)--(0.0301209,1.82645)--(0.0325181,1.81967)--(0.034966,1.81301)--(0.037461,1.80645)--(0.04,1.8)--(0.04258,1.79365)--(0.0451981,1.7874)--(0.0478516,1.78125)--(0.0505378,1.77519)--(0.0532544,1.76923)--(0.0559991,1.76336)--(0.0587695,1.75758)--(0.0615637,1.75188)--(0.0643796,1.74627)--(0.0672154,1.74074)--(0.0700692,1.73529)--(0.0729394,1.72993)--(0.0758244,1.72464)--(0.0787226,1.71942)--(0.0816327,1.71429)--(0.0845531,1.70922)--(0.0874826,1.70423)--(0.0904201,1.6993)--(0.0933642,1.69444)--(0.0963139,1.68966)--(0.0992682,1.68493)--(0.102226,1.68027)--(0.105186,1.67568)--(0.108148,1.67114)--
(0.1111,5/3)--(0.111852,1.66892)--(0.112591,1.67123)--(0.113329,1.67361)--(0.114065,1.67606)--(0.114796,1.67857)--(0.115522,1.68116)--(0.116241,1.68382)--(0.116953,1.68657)--(0.117654,1.68939)--(0.118343,1.69231)--(0.119019,1.69531)--(0.119678,1.69841)--(0.120317,1.70161)--(0.120935,1.70492)--(0.121528,1.70833)--(0.122091,1.71186)--(0.122622,1.71552)--(0.123115,1.7193)--(0.123565,1.72321)--(0.123967,1.72727)--(0.124314,1.73148)--(0.1246,1.73585)--(0.124815,1.74038)--(0.124952,1.7451)--(0.125,1.75)--(0.124948,1.7551)--(0.124783,1.76042)--(0.124491,1.76596)--(0.124055,1.77174)--(0.123457,1.77778)--(0.122676,1.78409)--(0.121687,1.7907)--(0.120465,1.79762)--(0.118977,1.80488)--(0.117188,1.8125)--(0.115056,1.82051)--(0.112535,1.82895)--(0.109569,1.83784)--(0.106096,1.84722)--(0.102041,1.85714)--(0.0973183,1.86765)--(0.0918274,1.87879)--(0.0854492,1.89063)--(0.0780437,1.90323)--(0.0694444,1.91667)--(0.059453,1.93103)--(0.0478316,1.94643)--(0.0342936,1.96296)--(0.0184911,1.98077);

\draw[red!50!black,ultra thick] 
(0,2)--(0.0000980296,1.9901)--(0.000384468,1.98039)--(0.000848336,1.97087)--(0.00147929,1.96154)--(0.00226757,1.95238)--(0.00320399,1.9434)--(0.00427985,1.93458)--(0.00548697,1.92593)--(0.00681761,1.91743)--(0.00826446,1.90909)--(0.00982063,1.9009)--(0.0114796,1.89286)--(0.0132352,1.88496)--(0.0150816,1.87719)--(0.0170132,1.86957)--(0.019025,1.86207)--(0.0211118,1.8547)--(0.0232692,1.84746)--(0.0254925,1.84034)--(0.0277778,1.83333)--(0.0301209,1.82645)--(0.0325181,1.81967)--(0.034966,1.81301)--(0.037461,1.80645)--(0.04,1.8)--(0.04258,1.79365)--(0.0451981,1.7874)--(0.0478516,1.78125)--(0.0505378,1.77519)--(0.0532544,1.76923)--(0.0559991,1.76336)--(0.0587695,1.75758)--(0.0615637,1.75188)--(0.0643796,1.74627)--(0.0672154,1.74074)--(0.0700692,1.73529)--(0.0729394,1.72993)--(0.0758244,1.72464)--(0.0787226,1.71942)--(0.0816327,1.71429)--(0.0845531,1.70922)--(0.0874826,1.70423)--(0.0904201,1.6993)--(0.0933642,1.69444)--(0.0963139,1.68966)--(0.0992682,1.68493)--(0.102226,1.68027)--(0.105186,1.67568)--(0.108148,1.67114)--(0.1111,5/3);

\filldraw[green!30!black, thick]
(0.1111,5/3)--(0.111852,1.66892)--(0.112591,1.67123)--(0.113329,1.67361)--(0.114065,1.67606)--(0.114796,1.67857)--(0.115522,1.68116)--(0.116241,1.68382)--(0.116953,1.68657)--(0.117654,1.68939)--(0.118343,1.69231)--(0.119019,1.69531)--(0.119678,1.69841)--(0.120317,1.70161)--(0.120935,1.70492)--(0.121528,1.70833)--(0.122091,1.71186)--(0.122622,1.71552)--(0.123115,1.7193)--(0.123565,1.72321)--(0.123967,1.72727)--(0.124314,1.73148)--(0.1246,1.73585)--(0.124815,1.74038)--(0.124952,1.7451)--(0.125,1.75)--(0.124948,1.7551)--(0.124783,1.76042)--(0.124491,1.76596)--(0.124055,1.77174)--(0.123457,1.77778)--(0.122676,1.78409)--(0.121687,1.7907)--(0.120465,1.79762)--(0.118977,1.80488)--(0.117188,1.8125)--(0.115056,1.82051)--(0.112535,1.82895)--(0.1111,1.8325)--(0.1111,5/3);

\filldraw[black] (0.1111,5/3) circle (0.15pt) node [below = 2pt, align=center] {\large Cheap Talk};

\node [align=center] at (-0.1,5/3+0.2) {\large Voluntary};
\node [align=center] at (-0.1,5/3+0.17) {\large Disclosure};
\draw [-{Stealth[length=2mm, width=1mm]}] (-0.03,5/3+0.16) -- (0.03,5/3+0.14);

\node [align=center] at (0.26,5/3+0.2) {\large Request};
\node [align=center] at (0.26,5/3+0.17) {\large Consent};
\draw [-{Stealth[length=2mm, width=1mm]}] (0.2,5/3+0.16) -- (0.13,5/3+0.11);



\draw[black] (0.07,1.85) node [below = 2pt, align=center] {\large \eqmset};
\end{tikzpicture}
    }
    \caption{Equilibrium outcomes selected by different communication protocols. Seller types are uniformly distributed on $[0,1]$, and $(L, H) = (1,3)$.}\label{fig:uniform-protocols}
\end{figure}

Below, we describe protocols that select different subsets of the implementable outcomes, as depicted in \autoref{fig:uniform-protocols}.
For simplicity, assume that the seller's type distribution $\typemeas$ has a full support on $[0,1]$.
\paragraph{Cheap Talk.} As a benchmark, suppose that the buyer knows the value and can inform the seller about it via an unverifiable, cheap-talk message---i.e., $\typeb\equiv\wtp$, $\ass=\{a_0\}$, $\asb(\high) = \asb(\low)=\asb$, and $s\equiv a_B$.
This protocol leads to a no-information outcome.
Indeed, if an equilibrium entailed some nontrivial communication---i.e.,  if $\wtp = \low$  were strictly more likely after message $m\in\asb$ than after message $m'\in\asb$---then sending $m$ would lead to lower prices for all seller types and strictly so, given the full-support assumption.
An $\high$-buyer would then never send message $m'$, which would contradict the presumption that message $m$ indicates a higher likelihood of an $L$-buyer.
The result implies that for informative communication to occur, information provision must be verifiable to some extent.

\paragraph{Voluntary Disclosure.} 
We turn to the buyer's voluntary disclosure of verifiable information.\footnote{If the seller is uninformed, this setting is a special case of \citet*{ali2020voluntary}.}
In this protocol, given the full-support assumption, the buyer is never willing to verify that he has value $\high$, because the buyer would then face a high price and lose a positive surplus.
As a result, any equilibrium strategy of the buyer generates a low-value flagging signal.

Formally, we consider the following protocol: $\typeb\equiv \wtp$, $\ass=\{a_0\}$, $\asb(\wtp) = \{ \{v\}, \emptyset\}$ for each $\wtp \in \{\high, \low\}$, and $s\equiv \ab$, so that the buyer can choose whether to disclose his value.
We argue that the set of equilibrium outcomes coincides with all implementable outcomes  spanned by the low-value flagging signals.
Indeed, if an $\low$-buyer sends message $\emptyset$ with a positive probability, then the seller with a sufficiently low type $\type$ sets price $\low$ after observing message $\emptyset$.
But then, an $\high$-buyer strictly prefers sending $\emptyset$ to $\{\high\}$.
Thus, there could be only two kinds of equilibrium: (i) an $\low$-buyer mixes between messages $\{\low\}$ and $\emptyset$ and an $\high$-buyer sends $\emptyset$ with probability $1$, and (ii) an $\low$-buyer sends message $\{\low\}$ for sure, and an $\high$-buyer sends $\emptyset$ or $\{\high\}$ revealing his value $\high$.
In fact, any such buyer strategy can arise in equilibrium, because an $\low$-buyer never obtains a positive surplus and thus is indifferent between any messages, and any possible deviation by an $\high$-buyer can be deterred by the seller's skeptical belief that places probability $1$ on $\high$.

The low-value flagging characterization, combined with \autoref{propositionUniform}, implies that if the seller's type is uniformly distributed, the equilibrium outcomes of the voluntary disclosure game  span the left boundary of the surplus set (see \autoref{fig:uniform}), which highlights the potential inefficiency of this protocol. This inefficiency remains even if  the seller can communicate  before the buyer's disclosure---i.e., if $\ass$ is general. 
In that case, in equilibrium, each message sent by the seller induces some low-value flagging signal, but any two low-value flagging signals are ranked in Blackwell informativeness.
Hence, all seller types send a message that induces the same low-value flagging signal.

\paragraph{Request-Consent Protocol.}
The seller's action space $A_S$ is given by the set of all signals with $S=[0,1]$, and the buyer's action space is given by $A_B(\high) = A_B(\low) = \{accept, reject\}$.
    The seller first chooses a signal, and then the buyer observes the requested signal (but not its realization) and decides whether to accept it.
The seller obtains the requested signal if the buyer chooses $accept$ and does not observe any additional information otherwise.

In this protocol, if the buyer is perfectly informed (i.e., $t\equiv \wtp$), then any implementable outcome can arise in some equilibrium:
Take any implementable outcome and the public signal $\info$ that implements it.
The following equilibrium has all seller types obtain signal $\info$.
On the equilibrium path, all seller types request signal $\info$ and the buyer accepts it regardless of his value.
The seller's deviation to an off-path signal can be deterred if the buyer, following the deviation, believes that the seller's type is $\type =0$.
The buyer with this belief thinks that the seller will set price $\low$, so the buyer never strictly benefits from providing additional information.
In turn, if the seller believes that any buyer who deviates and rejects the signal request has value $\high$, then the buyer finds it optimal to accept signal $\info$.

In contrast, if the buyer is not informed when deciding on the request acceptance (i.e., $t\equiv{t_0}$), then the set of equilibrium outcomes is smaller and coincides with the set of implementable outcomes such that the buyer is at least as well off as under no data provision.
Indeed, if all seller types request the same signal $\info$, the uninformed buyer will accept it if and  only if signal $\info$ weakly increases his ex ante payoff; the seller cannot use the off-path belief punishment because the buyer is uninformed.  The result then follows from \autoref{prop:public}.
Importantly, in some cases  there may be no public signal that strictly increases buyer surplus (see \autoref{fig:uniform} (b)).
In those cases, the unique equilibrium outcome of the request-consent protocol with an uninformed buyer would be no data collection.

\newpage
\subsection{Many Values}
There are multiple ways in which our setting can be extended. 
One is to allow many possible buyer values, $\wtps = \{\wtp_1,..., \wtp_n\}$ with $v_1<\dots<v_n$, $n>2$. The seller's type $\type \in \Delta(\wtps)$ is now multidimensional, and the complete characterization of implementable outcomes is not attainable because it involves multidimensional screening. Nevertheless, we show that our main qualitative findings extend to this setting and provide necessary clarifications.

First of all, private information held by the seller generally limits the set of implementable outcomes, as in the case of binary values. However, in the case of many values, the limits are less stark, since public signals do not necessarily span all implementable outcomes. Differentiated data provision can be welfare enhancing in this setting, and it can even attain the buyer's first-best outcome, which attains efficiency without increasing the seller's profits.
The following example illustrates such a possibility.

\begin{example}(Benefits of Heterogeneous Data Provision)
Let the possible values be $\wtps=\{\low,\med,\high\}=\{1,3,4\}$. Let there be three seller types, $\type_1=(1/2,1/4,1/4)$, $\type_2=(1/2,1/2,0)$, and $\type_3=(1/2,0,1/2)$, where the three types are equally likely.
In the absence of additional data, the types set prices $p_1=\med$, $p_2=\med$, and $p_3=\high$, and buyer surplus is $\cs_0=1/12$.

We derive the buyer-optimal outcome and show that it is efficient and cannot be attained by a public signal.
Namely, consider a menu in which the types are provided with the following signals:
\begin{align}
    \begin{array}{c|cc}
    \info(\type_1) & s_{1} & s_{2}\\
    \hline 
    \low & 1 & 0\\
    \med & 1/2 & 1/2\\
    \high & 1/2 & 1/2
    \end{array},\quad\quad
    \begin{array}{c|cc}
    \info(\type_2) & s_{1} & s_{2}\\
    \hline 
    \low & 1 & 0\\
    \med & 1/2 & 1/2\\
    \high & 1/2 & 1/2
    \end{array},\quad\quad
    \begin{array}{c|cc}
    \info(\type_3) & s_{1} & s_{2}\\
    \hline 
    \low & 1 & 0\\
    \med & 2/3 & 1/3\\
    \high & 1/3 & 2/3
    \end{array},
\end{align}
so that each signal has two possible realizations with the likelihood function \map\ as presented in the tabular form. 
It is straightforward to verify that no type wants to deviate.
Moreover, the outcome is efficient and each type earns the same profits as in the absence of additional information: Given these signals, types $\type_1$ and $\type_2$ set prices $p=L$ and $p=M$ after signal realizations $s_1$ and $s_2$, respectively, and type $\type_3$ sets prices $p=L$ and $p=H$, respectively.
The buyer surplus is then maximal among all feasible allocations, given the individual rationality of the seller.

However, we cannot implement this outcome by providing the same signal to all seller types:
Any efficient public signal must give strictly positive rents to type $\type_1$. 
Indeed, for signal $\info = (\signals, \map)$ to attain efficiency, it must be that whenever $\pi(s_0|\low)>0$ for some $s_0\in\signals$ all types charge $\price=\low$ following that signal, which means that $\pi(s_0|\low)\geq2\pi(s_0|\med)$ to persuade type $\type_2$ and $\pi(s_0|\low)\geq3\pi(s_0|\high)$ to persuade type $\type_3$; in turn, these incentive constraints together guarantee that $\type_1$ also charges a low price. Since type $\type_1$ never charges a price below $\price=\low$, her profit decreases as we increase the frequency of signal $s_0$---i.e., as we increase $\pi(s_0|\low)$, $\pi(s_0|\med)$, and $\pi(s_0|\high)$ subject to the incentive constraints. 
Therefore, of all efficient public signals, the profit of type $\type_1$ is minimized at the signal
\begin{align}
    \begin{array}{c|cc}
    \hat\info & s_{1} & s_{2}\\
    \hline 
    \low & 1 & 0\\
    \med & 1/2 & 1/2\\
    \high & 1/3 & 2/3
    \end{array},
\end{align}
which is efficient and maximizes the probability of type $\type_1$ setting price $p=L$ while ensuring that type $\type_1$ offers the remaining buyers the second-lowest price $p=\med$. 
The resulting $\type_1$'s minimal profit  equals $19/12$, which is strictly greater than her outside option of $3/2$ and, as such, necessarily results in lower buyer surplus than under the multi-item mechanism presented above.\hfill$\square$
\end{example}

Intuitively, as in most mechanism-design problems, the multidimensionality of the seller's type gives rise to a richer structure of incentive constraints and implementable outcomes. In particular, the buyer-optimal efficient signals are not necessarily Blackwell-comparable and can appeal to the ``right" buyer type. As a result, complete characterization of the set of implementable outcomes becomes more complicated. Nevertheless, we  show that the welfare implications of \autoref{sec:welfare-efficiency} generalize: Seller heterogeneity introduces a trade-off between efficiency and consumer welfare.\footnote{
All conditions of \autoref{corollaryInefficiencyMany}  hold, for example, whenever $F$ has full support over $\Delta(V)$.}


\begin{proposition}\emph{(Efficiency and Buyer Surplus)} \label{corollaryInefficiencyMany}
Assume that the type distribution $F$ places probability 1 on the interior of $\Delta(V)$. Then the following hold:
\begin{enumerate}[leftmargin=0.55cm]
\item  If $F$ admits a positive density over an open set of types who, in the absence of data, charge prices $p>v_1$, then any efficient outcome gives a strictly positive rent to the seller. 
\item If there is a strictly positive measure of types in any neighborhood of the type that places probability 1 on $v_n$, then the only efficient outcome is that under full information.
\end{enumerate}
\end{proposition}
\autoref{corollaryInefficiencyMany} is a counterpart of \autoref{corollaryInefficiency} for the case of many values. The first part of \autoref{corollaryInefficiencyMany} is based on the fact that signals that yield no rents to the seller must be carefully tailored to the seller's type. Specifically, consider the types over which $F$ has a density. Efficiency requires that these types occasionally set price $p=v_1$ to serve the lowest-value buyer, and therefore must be given additional information. However, if some type weakly prefers to set the lowest price after receiving some signal realization, nearby types would strictly prefer to set the lowest price following that realization and would therefore strictly prefer to change their prior action. 
This means that they can earn strictly positive rents from that signal and must earn strictly positive rents when faced with any efficient menu.

For the second part of \autoref{corollaryInefficiencyMany}, we can show that under the stated condition, there exists a sequence of types with the following properites.
First, types converge to the extreme type that places probability $1$ on the highest possible value.
Second, along the sequence, the corresponding signals, which induce efficient outcomes, converge to the fully informative signal. In particular, we show that the types along this sequence need to be persuaded to charge all possible prices  and thus must be provided with progressively more detailed information. The limit argument, coupled with incentive compatibility, then implies that all types must be offered a fully informative signal.


\section{Conclusion}

We studied the provision of consumer data to a monopoly seller who has imperfect private information
about the value of its product. Without private information, the designer can flexibly provide the seller with data to distribute efficient total surplus between the buyer and the seller. We show that the seller’s private information prevents the designer from providing different signals to different seller types, which leads to the impossibility of effective screening. Our results provide insights into data policies for policymakers and businesses. For policymakers or platforms that aim to control the follow of consumer data to sellers, the results highlight the trade-off between consumer protection and efficiency driven by adverse selection regarding data use. From the seller’s perspective, our results clarify why the collection of third-party data without consumer consent might harm profits. To clarify the economic intuition, we have focused on a simple setting in which the buyer has binary values and the seller uses data for price discrimination. However, the issues that arise when a firm's private information hinders the effective allocation of consumer data---and the resulting challenges faced by regulators and firms with regard to data policies---would be relevant in broader contexts.

\bibliography{privacy}

\begin{thebibliography}{29}
\newcommand{\enquote}[1]{``#1''}
\expandafter\ifx\csname natexlab\endcsname\relax\def\natexlab#1{#1}\fi

\bibitem[\protect\citeauthoryear{Acquisti, Taylor, and Wagman}{Acquisti
  et~al.}{2016}]{acquisti2016economics}
\textsc{Acquisti, A., C.~Taylor, and L.~Wagman} (2016): \enquote{The Economics
  of Privacy,} \emph{Journal of Economic Literature}, 54, 442--492.

\bibitem[\protect\citeauthoryear{Ali, Lewis, and Vasserman}{Ali
  et~al.}{forthcoming}]{ali2020voluntary}
\textsc{Ali, S.~N., G.~Lewis, and S.~Vasserman} (forthcoming):
  \enquote{Voluntary Disclosure and Personalized Pricing,} \emph{Review of
  Economic Studies}.

\bibitem[\protect\citeauthoryear{Alonso and Camara}{Alonso and
  Camara}{2016}]{alca16}
\textsc{Alonso, R. and O.~Camara} (2016): \enquote{Bayesian Persuasion with
  Heterogeneous Priors,} \emph{Journal of Economic Theory}, 165, 672--706.

\bibitem[\protect\citeauthoryear{Alonso and Zachariadis}{Alonso and
  Zachariadis}{2021}]{alza21}
\textsc{Alonso, R. and K.~Zachariadis} (2021): \enquote{Persuading Large
  Investors,} \emph{Working paper}.

\bibitem[\protect\citeauthoryear{Argenziano and Bonatti}{Argenziano and
  Bonatti}{2021}]{argenziano2020information}
\textsc{Argenziano, R. and A.~Bonatti} (2021): \enquote{Information Revelation
  and Privacy Protection,} \emph{Working paper}.

\bibitem[\protect\citeauthoryear{Aumann and Maschler}{Aumann and
  Maschler}{1995}]{auma95}
\textsc{Aumann, R. and M.~Maschler} (1995): \emph{Repeated Games with
  Incomplete Information}, MIT Press.

\bibitem[\protect\citeauthoryear{Baron and Myerson}{Baron and
  Myerson}{1982}]{baron1982regulating}
\textsc{Baron, D.~P. and R.~B. Myerson} (1982): \enquote{Regulating a
  Monopolist with Unknown Costs,} \emph{Econometrica}, 911--930.

\bibitem[\protect\citeauthoryear{Bergemann, Bonatti, and Gan}{Bergemann
  et~al.}{2022}]{bergemann2019economics}
\textsc{Bergemann, D., A.~Bonatti, and T.~Gan} (2022): \enquote{The Economics
  of Social Data,} \emph{The RAND Journal of Economics}, 53, 263--296.

\bibitem[\protect\citeauthoryear{Bergemann, Bonatti, and Smolin}{Bergemann
  et~al.}{2018}]{bergemann2018design}
\textsc{Bergemann, D., A.~Bonatti, and A.~Smolin} (2018): \enquote{The Design
  and Price of Information,} \emph{American Economic Review}, 108, 1--48.

\bibitem[\protect\citeauthoryear{Bergemann, Brooks, and Morris}{Bergemann
  et~al.}{2015}]{bergemann2015limits}
\textsc{Bergemann, D., B.~Brooks, and S.~Morris} (2015): \enquote{The Limits of
  Price Discrimination,} \emph{American Economic Review}, 105, 921--957.

\bibitem[\protect\citeauthoryear{Choi, Jeon, and Kim}{Choi
  et~al.}{2019}]{choi2019privacy}
\textsc{Choi, J.~P., D.-S. Jeon, and B.-C. Kim} (2019): \enquote{Privacy and
  Personal Data Collection with Information Externalities,} \emph{Journal of
  Public Economics}, 173, 113--124.

\bibitem[\protect\citeauthoryear{Condorelli and Szentes}{Condorelli and
  Szentes}{2022}]{cosz22}
\textsc{Condorelli, D. and B.~Szentes} (2022): \enquote{Buyer-Optimal Platform
  Design,} \emph{Working paper}.

\bibitem[\protect\citeauthoryear{Deb and Roesler}{Deb and
  Roesler}{2022}]{roesler2022multi}
\textsc{Deb, R. and A.-K. Roesler} (2022): \enquote{Multi-dimensional
  Screening: Buyer-Optimal Learning and Informational Robustness,}
  \emph{Working paper}.

\bibitem[\protect\citeauthoryear{Doval and Smolin}{Doval and
  Smolin}{2023}]{dosm21}
\textsc{Doval, L. and A.~Smolin} (2023): \enquote{Persuasion and Welfare,}
  \emph{Working paper}.

\bibitem[\protect\citeauthoryear{Elliott, Galeotti, Koh, and Li}{Elliott
  et~al.}{2022}]{elliott2021market}
\textsc{Elliott, M., A.~Galeotti, A.~Koh, and W.~Li} (2022): \enquote{Market
  Segmentation through Information,} \emph{Working paper}.

\bibitem[\protect\citeauthoryear{Fainmesser, Galeotti, and Momot}{Fainmesser
  et~al.}{2022}]{fainmesser2022digital}
\textsc{Fainmesser, I.~P., A.~Galeotti, and R.~Momot} (2022): \enquote{Digital
  privacy,} \emph{Management Science}.

\bibitem[\protect\citeauthoryear{Haghpanah and Siegel}{Haghpanah and
  Siegel}{2022}]{haghpanah2022limits}
\textsc{Haghpanah, N. and R.~Siegel} (2022): \enquote{The Limits of
  Multiproduct Price Discrimination,} \emph{American Economic Review:
  Insights}, 4, 443--58.

\bibitem[\protect\citeauthoryear{Haghpanah and Siegel}{Haghpanah and
  Siegel}{forthcoming}]{haghpanah2019pareto}
---\hspace{-.1pt}---\hspace{-.1pt}--- (forthcoming): \enquote{Pareto Improving
  Segmentation of Multi-product Markets,} \emph{Journal of Political Economy}.

\bibitem[\protect\citeauthoryear{Kamenica and Gentzkow}{Kamenica and
  Gentzkow}{2011}]{kage11}
\textsc{Kamenica, E. and M.~Gentzkow} (2011): \enquote{Bayesian Persuasion,}
  \emph{American Economic Review}, 101, 2590--2615.

\bibitem[\protect\citeauthoryear{Kolotilin, Mylovanov, Zapechelnyuk, and
  Li}{Kolotilin et~al.}{2017}]{kolotilin2017persuasion}
\textsc{Kolotilin, A., T.~Mylovanov, A.~Zapechelnyuk, and M.~Li} (2017):
  \enquote{Persuasion of a Privately Informed Receiver,} \emph{Econometrica},
  85, 1949--1964.

\bibitem[\protect\citeauthoryear{Myerson}{Myerson}{1982}]{myerson1982optimal}
\textsc{Myerson, R.~B.} (1982): \enquote{Optimal Coordination Mechanisms in
  Generalized Principal--Agent Problems,} \emph{Journal of Mathematical
  Economics}, 10, 67--81.

\bibitem[\protect\citeauthoryear{Myerson}{Myerson}{1983}]{myerson1983mechanism}
---\hspace{-.1pt}---\hspace{-.1pt}--- (1983): \enquote{Mechanism Design by an
  Informed Principal,} \emph{Econometrica}, 1767--1797.

\bibitem[\protect\citeauthoryear{Rayo and Segal}{Rayo and
  Segal}{2010}]{rayo2010optimal}
\textsc{Rayo, L. and I.~Segal} (2010): \enquote{Optimal Information
  Disclosure,} \emph{Journal of Political Economy}, 118, 949--987.

\bibitem[\protect\citeauthoryear{Rhodes and Zhou}{Rhodes and
  Zhou}{2022}]{rhodes2022personalized}
\textsc{Rhodes, A. and J.~Zhou} (2022): \enquote{Personalized Pricing and
  Competition,} \emph{Working paper}.

\bibitem[\protect\citeauthoryear{Roesler and Szentes}{Roesler and
  Szentes}{2017}]{roesler2017buyer}
\textsc{Roesler, A.-K. and B.~Szentes} (2017): \enquote{Buyer-Optimal Learning
  and Monopoly Pricing,} \emph{American Economic Review}, 107, 2072--80.

\bibitem[\protect\citeauthoryear{Shaked and Shanthikumar}{Shaked and
  Shanthikumar}{2007}]{shaked2007stochastic}
\textsc{Shaked, M. and J.~G. Shanthikumar} (2007): \emph{Stochastic Orders},
  Springer.

\bibitem[\protect\citeauthoryear{Shi and Zhang}{Shi and
  Zhang}{2020}]{shi2020welfare}
\textsc{Shi, X. and J.~Zhang} (2020): \enquote{Welfare of Price Discrimination
  and Market Segmentation in Duopoly,} \emph{Working paper}.

\bibitem[\protect\citeauthoryear{Smolin}{Smolin}{forthcoming}]{smolin2020disclosure}
\textsc{Smolin, A.} (forthcoming): \enquote{Disclosure and Pricing of
  Attributes,} \emph{The RAND Journal of Economics}.

\bibitem[\protect\citeauthoryear{Yang}{Yang}{2022}]{yang2022selling}
\textsc{Yang, K.~H.} (2022): \enquote{Selling Consumer Data for Profit: Optimal
  Market-Segmentation Design and Its Consequences,} \emph{American Economic
  Review}, 112, 1364--93.

\end{thebibliography}
\bibliographystyle{ecta}

\section*{\hfil Appendix: Proofs Omitted from the Main Text \hfil}\label{appendix}

\renewcommand\thesubsection{\Alph{subsection}}


\begin{proof}[Proof of \autoref{claim:observable}]
To show the ``only if" direction, take any feasible welfare outcome 
\begin{equation*}
	(\cs, \profit) = \left(\int^1_0 \cs(\type) \dint F(\type), \int^1_0 \profit(\type) \dint F(\type) \right).
\end{equation*}
\hyperref[claim:bbm]{\autoref{claim:bbm}} implies that for each $\type$, we have $\cs(\type) \ge 0$, $\profit(\type) \ge \psmin(\type)$, and $\cs(\type) + \profit(\type) \le \tsmax(\type)$.
Integrating both sides of each inequality with $F$, we obtain $\cs \ge 0$, $\profit \ge \psmin$, and $\cs + \profit \le \tsmax$.

To show the ``if" direction, take any $(\cs, \profit) \in \R^2$ such that $\cs \ge 0$, $\profit \ge \psmin$, and $\cs + \profit \le \tsmax$.
The point $(\cs, \profit)$ belongs to the triangle whose vertices are $(0, \tsmax)$, $(0, \psmin)$, and $(\tsmax - \psmin, \psmin)$.
Let $\alpha, \beta \in [0,1]$ with $\alpha + \beta \le 1$ satisfy 
\begin{align*}
	(\cs, \profit) &= \alpha (0, \tsmax) + \beta (0, \psmin) + (1 - \alpha - \beta)(\tsmax - \psmin, \psmin)\\
		       &=\left(  \int^1_0  (1 - \alpha - \beta) [\tsmax(\type)-\psmin(\type)] \dint F(\type), 
			       \int^1_0  \alpha \tsmax(\type) + \beta \psmin(\type) + (1 - \alpha - \beta)\psmin(\type)\dint F(\type)\right).
\end{align*}
For each $\type$, the point 
\begin{equation*}
	(\cs(\type), \profit(\type)) \triangleq \Big( (1 - \alpha - \beta) [\tsmax(\type)-\psmin(\type)] , \alpha \tsmax(\type) + \beta \psmin(\type) + (1 - \alpha - \beta)\psmin(\type)\Big)
\end{equation*}
	is in the surplus triangle and thus feasible.
Aggregating the welfare outcome $(\cs(\type), \profit(\type))$ across possible $\type$, we conclude that
$(\cs, \profit)  = \left(\int^1_0 \cs(\type)\dint F(\type), \profit(\type)\dint F(\type)\right)$ is also feasible.
\end{proof}

\begin{proof}[Proof of \autoref{prop:structural}]\label{sectionAppendixBinary}
Part 1. Fix any $\type_2>\type_1$.  The system of mutual  incentive constraints is 
\begin{align}
(1-\alpha(\type_1))L+\type_1(\alpha(\type_1) L + \beta(\type_1) (H-L)) \geq (1-\alpha(\type_2))L+\type_1(\alpha(\type_2) L + \beta(\type_2) (H-L)), \label{eq:ic_two_types_12}\\
(1-\alpha(\type_2))L+\type_2(\alpha(\type_2) L + \beta(\type_2) (H-L)) \geq (1-\alpha(\type_1))L+\type_2(\alpha(\type_1) L + \beta(\type_1) (H-L)).
\label{eq:ic_two_types_21}
\end{align}
Summing over the inequalities (\ref{eq:ic_two_types_12}) and (\ref{eq:ic_two_types_21}) and using the fact that $\theta_2>\theta_1$ we obtain
\begin{align} \label{eq:binary_type_mon_1}
    \alpha(\type_2) L + \beta(\type_2) (H-L)\geq  \alpha(\type_1) L + \beta(\type_1) (H-L).
\end{align}
In turn,  (\ref{eq:ic_two_types_12}) and (\ref{eq:binary_type_mon_1}) together imply $\alpha(\type_2)\geq\alpha(\type_1)$ because 
\begin{align}\label{eq:binary_type_mon_2}
    (\alpha(\type_2)-\alpha(\type_1))L\geq\type_1(\alpha(\type_2) L + \beta(\type_2) (H-L)-\alpha(\type_1) L - \beta(\type_1) (H-L))\geq0.
\end{align}
Finally,  (\ref{eq:ic_two_types_21}) and (\ref{eq:binary_type_mon_2}) imply $\beta(\type_2)\geq\beta(\type_1)$ because 
\begin{align}
    (\beta(\type_2)-\beta(\type_1))\type_2(H-L)\geq (\alpha(\type_2)-\alpha(\type_1))(1-\type_2)L\geq0.
\end{align}

Part 2. Fix any $\type_1<\type_2<\type_3$. The system of incentive constraints of type $\type_2$ toward types $\type_1$ and $\type_3$ can be written as
\begin{align}
(\beta(\type_2)-\beta(\type_1))\type_2(H-L)\geq (\alpha(\type_2)-\alpha(\type_1))(1-\type_2)L, \label{eq:ic_three_types_21}\\
(\beta(\type_3)-\beta(\type_2))\type_2(H-L)\leq (\alpha(\type_3)-\alpha(\type_2))(1-\type_2)L.
\label{eq:ic_three_types_23}
\end{align}
By property in Part 1, all sides of (\ref{eq:ic_three_types_21}) and (\ref{eq:ic_three_types_23}) are positive. Multiplying the respective smaller and larger parts and dividing the resulting inequality by $\theta_2(1-\theta_2)(\high -\low)\low$, 
we obtain the desired inequality of Part 2.
\end{proof}

\begin{proof}[Proof of \autoref{propositionUniform}] \textit{Step 1.} We use the ``boundary" to mean the boundary of $\eqmset$. By \autoref{prop:extreme_outcomes} and \autoref{cor:signal_complexity}, any extreme point on the boundary can arise with a signal that has two signal realizations.
We parameterize such signals by $(\alpha,\beta)$, $\beta \ge \alpha$ as
\begin{align}
    \begin{array}{c|ccc}
    \info(\type) & s_{L} & s_{H} \\
    \hline v=L & 1-\alpha & \alpha\\
    v=H & 1-\beta  & \beta
    \end{array}.
\end{align}
 Given any such signal, we write the objective $\lambda_{U} \cs + \lambda_\profit \profit$ in terms of the highest and lowest types that respond to the signal realizations. In particular, for any given $(\alpha, \beta)$ with $\beta \ge \alpha$, we can find two cutoffs $x,y$ with $x \le y$ such that types below $x$ set price $\low$ after both realizations; types above $y$ set price $\high$ after both realizations; and types between $x$ and $y$ will set prices $\high$ and $\low$ after $s_H$ and $s_L$, respectively. 
Cutoffs $x$ and $y$ solve  
\begin{align*}
	L &= (1-\alpha)L + x\left(\alpha L+\beta (H-L)\right),\\
    y    H &=   (1-\alpha)L+y \left(\alpha L+\beta (H-L)\right), 
\end{align*}
which implies
\begin{align*}
	 x = \frac{\alpha L}{\alpha L+\beta (H-L)},\quad 
	 y = \frac{(1-\alpha)L}{H-\alpha L-\beta (H-L)}.
\end{align*}  
Alternatively, we can write $(\alpha, \beta)$ as functions of $(x,y)$:
 \begin{align*}
	 \alpha = \frac{ x(Hy-L)}{L(y-x)},\quad
	 \beta = \frac{(Hy-L)(1-x)}{(y-x)(H-L)}.
  \end{align*} 
Cutoffs $(x,y)$ can arise under some signal if and only if
$0\le x \le \frac{\low}{\high} \le y\le 1$.
The interim profit of type $\type \in [x,y]$ is 
\begin{equation*}
	\profit(x,y\mid \theta)= (1-\type)(1-\alpha)L+ \type( (1 - \beta)L +  \beta H)) = L + \frac{(Hy-L)(\type-x)}{y-x}.
\end{equation*}
The interim profits of types $\type \le x$ and $\type \ge y$ are $\low$ and $\type \high$, respectively.
The ex ante seller profit is 
\begin{align}
		 \profit(x,y) &= \int^x_0  \low \dint\type + \int^y_x \low + \frac{\high y - \low}{y-x}(\type - x) \dint\type + \int^1_y \type \high \dint\type\notag\\
	&=
	Ly + \frac{1}{2}(\high y - \low) (y+x) - (\high y - \low) x+ \frac{\high}{2}(1-y^2).
\end{align}  
The buyer surplus is 
\begin{align}
\cs(x,y)\triangleq
&(\high - \low )\int^x_0  \type \dint\type +(1-\beta)(H-L)\int^y_x \type  \dint\type\notag\\
=&(\high - \low )\int^x_0  \type \dint\type +\left(\high -\low-\frac{(\high y - \low)(1-x)}{y - x}\right) \int^y_x \type  \dint\type\notag\\
	=&(\high - \low )\int^y_0  \type \dint\type -\frac{(\high y - \low)(1-x)}{y - x} \int^y_x \type  \dint\type\notag\\
	=&\frac{1}{2}(\high - \low )y^2  -\frac{1}{2}(\high y - \low)(1-x)(y + x).
\end{align} 

\medskip

\noindent {\textit{Step 2.}}
We characterize signals that span the ``right'' boundary that corresponds to $\lambda_{U}  \ge 0$.
Take any $(\lambda_{U}, \lambda_\profit) \in \R^2$ with $\lambda_{U} \ge 0$.
Because $\profit(x,y)$ is linear in $x$ and $\cs(x,y)$ is convex in $x$, $W(x,y)$ is convex in $x$ and maximized at $x=0$ or $x=\low/\high$.
Note that $(x,y) =(\low/\high, y)$ implies $(\alpha, \beta) = (1,1)$ and  $(x,y) =(0, \low/\high)$ leads to $(0,0)$, but both are the same uninformative signal.
Thus we can without loss of generality assume $x=0$ and focus on the problem $\max_{\frac{\low}{\high}\le y\le 1} W(0,y)$, where
\begin{align*}
	 W(0,y) &= 
	 \lambda_{U} \left[\frac{1}{2}(\high - \low )y^2  -\frac{1}{2}(\high y - \low)y\right]+\lambda_\profit \left[Ly + \frac{1}{2}(\high y - \low) y + \frac{\high}{2}(1-y^2)\right]\\
		&=\frac{\lambda_{U}}{2}Ly (1-y)+\frac{\lambda_\profit}{2} (Ly + H).
\end{align*}

If $\lambda_{U} =0$, then the function is maximized at $y=1$ for $\lambda_\profit>0$ and at $y=0$ for $\lambda_\profit<0$.
Thus, the point that maximizes the seller profit is attained by the fully informative signal, and the point that minimizes the seller profit is attained by the uninformative signal.

If $\lambda_U >0$, the function $W(0,y)$ is strictly concave in $y$ so we can use the first-order condition to determine an interior solution: 
 \begin{align*}
	 y = \frac{\lambda_{U} + \lambda_\profit}{2\lambda_{U}}
	  =\frac{1}{2} \left(1 + \frac{\lambda_\profit}{\lambda_{U}} \right).
\end{align*}  
If $\frac{\low}{\high}\le \frac{1}{2} \left(1 + \frac{\lambda_\profit}{\lambda_{U}} \right) \le 1$, then 
the optimal $y$ is $\frac{1}{2} \left(1 + \frac{\lambda_\profit}{\lambda_{U}} \right)$.
Otherwise, there is a corner solution $y=\frac{\low}{\high}$ or $y=1$ for low or large values of $\lambda_\profit$, respectively.

Plugging $x=0$ into $(\alpha, \beta)$ above, we obtain $\alpha =0$---i.e., the signal sends realization $s_H$ only if the value is $\high$, so it is a high-value flagging signal.
Thus each point on the right boundary arises under some high-value flagging signal.
As we move the right boundary from the seller-worst point to the seller-optimal point, cutoff $y$ increases (or equivalently, $\beta$ increases) and the corresponding signal changes from the uninformative signal to the fully informative signal.
If $\frac{\low}{\high} \ge \frac{1}{2}$, the buyer optimal point is $\lambda_\profit=0$, so we have $y^* = \low/\high$---i.e., the uninformative signal maximizes the buyer surplus.
If $\frac{\low}{\high} < \frac{1}{2}$, the buyer optimal point is $y^* = 1/2$---i.e., a partially informative high-value flagging signal maximizes buyer surplus.
In this case, plugging the optimal $(x,y)$ into $(\alpha, \beta)$, we obtain $\alpha = 1$ and $\beta = \frac{H-2L}{H-L}$.

\medskip

\noindent {\textit{Step 3.}} We characterize the ``left" boundary that corresponds to $\lambda_{U} <0$.
The seller profit $\profit(x,y)$ is linear in $y$ and the buyer surplus is
strictly concave in $y$.
Because  $\lambda_{U}<0$, the function $W(x,y)$ is strictly convex in $y$, which means that
the optimal $y$ will be $\frac{\low}{\high}$ or $1$.
Because $y= \frac{\low}{\high}$ is equivalent to $(x,y) = (\frac{\low}{\high}, 1)$, 
we can without loss of generality assume that $y = 1$. 
We have 
\begin{align*}
	W(x,1)&=  \lambda_{U}\left[   \frac{1}{2}(\high - \low )  -\frac{1}{2}(\high  - \low)(1-x)(1 + x)\right]+\lambda_\profit\left[	  L + \frac{1}{2}(\high  - \low) (1+x) - (\high  - \low) x \right]\\
	 &=\frac{\lambda_{U}}{2}(\high - \low ) x^2+\frac{\lambda_\profit}{2}\left[    2     L + (\high  - \low) (1-x)  \right].
\end{align*}                                         
The first-order condition with respect to $x$ yields 
\begin{align}
	 \lambda_{U} x  - \frac{\lambda_\profit}{2} =0 \iff x = \frac{\lambda_\profit}{2\lambda_{U}}.
\end{align}
Thus any point on the left boundary can arise under a low-value flagging signal.
As we move the left boundary from the seller-worst point to the seller-optimal point by raising $\lambda_\profit$, the corresponding signal changes from no disclosure $(x,y) = (\low/\high, 1)$ to full disclosure $(x,y) = (0,1)$.
\end{proof}

\begin{proof}[Proof of \autoref{corollaryInefficiency}]
To prove part 1, take any efficient outcome. By \autoref{prop:public}, all types can be assumed to obtain the same signal.
If the signal is fully revealing, then the seller obtains a strictly positive rent because almost all seller types belong to the interior $(0,1)$ of the type space.
If the signal is not fully revealing, then there must exist  signal realization $s$ such that the posterior is non-degenerate and induces all types to set price $\low$. By the martingale property of belief, there must exist another signal realization $s'$ under which types in $(\low/\high, 1)$ optimally set price $\high$.
Because the outcome is efficient,   $s'$ perfectly reveals that the value is $\high$.
Then, there are two possibilities. First, if all types in $(\low/\high,1)$ are concentrated on a single point, then distribution $F$, which is non-degenerate, must place a positive probability on types in $[0, \low/\high]$. Those types would earn a strictly higher profit than without additional information, because they would strictly benefit from observing $s'$.
Second, if types strictly above $\low/\high$ are not concentrated on a single point, then below the highest type in the support of $F$ there is a positive measure of types who strictly prefer to set price $\low$ after observing signal realization $s$.
These types earn a strictly higher profit than without additional information.

To prove part 2, consider an implementable outcome in which a positive mass of types do not obtain the fully informative signal.
Profit $\profit(\type)$ of any such type satisfies $\profit(\type) < \type \high + (1- \type) \low$.
\autoref{prop:public} implies that there exists a public signal $\info$ that implements the same outcome.
Signal $\info$ is not fully informative, because otherwise we would have $\profit(\type) = \type \high + (1- \type) \low$ for all types.
Thus there is a set of signal realizations that can arise under both $\wtp = \high$ and $\wtp=\low$ with positive probabilities.
After observing such realizations, types that are sufficiently close to $1$  set price $\high$, which leads to no trade when $\wtp= \low$.
Because the set of any such types has a strictly positive measure, the efficiency loss is strictly positive.
\end{proof}

\begin{proof}[Proof of \autoref{prop:third-party-data}]
\textit{Part 1.}
If the seller's type is known to be $\type_0> \low/\high$, then in the absence of additional data the seller would set price $\high$,  leading to zero  buyer surplus.
As a result, at the constrained seller-optimal outcome the seller extracts full surplus.
In contrast, if the seller is privately informed, then the buyer surplus under no data provision is positive, because the seller  sets price $\low$ whenever her type is below $\low/\high$.
As a result, the buyer surplus under the constrained seller-optimal outcome is strictly positive, and
 $0=\cs < \cs'$, $\tsmax(\type_0)=\profit > \profit'$.

\textit{Part 2.} If the seller's type is known to be $\type_0 < \low/\high$, then in the absence of additional data the seller  sets price $\low$.
Because this is the uniquely best possible outcome for the buyer, the constrained seller-optimal outcome is the same as under no data provision.
In contrast, when the seller is privately informed, then she has a strictly positive measure of types above $\low/\high$ at which she sets price $\high$ and earns strictly higher profits than under no data provision.
Thus $\psmin(\type_0)=\profit < \profit'$, $\tsmax(\type_0)-\psmin(\type_0)=\cs > \cs'$. 
\end{proof}

\begin{proof}[Proof of \autoref{corollaryInefficiencyMany}]
Part 1. Assume the stated condition holds. Consider a direct menu that leads to an efficient outcome.
Let $\tilde\Theta\subseteq \Delta(V)$ be the open set over which $F$ has a positive density. 
By the condition described in Part 1, we can take $\tilde\Theta$ so that any type in $\tilde\Theta$ chooses a price strictly above the lowest possible value $v_1$ in the absence of additional information.
In this set, a positive measure of types assign positive probability to $v=v_1$, so there exists a type $\tilde\theta\in \tilde\Theta$ such that the direct signal $\info(\tilde\theta)$ recommends $p=v_1$ with a strictly positive probability. Let $\belief(\tilde\theta)$ be the posterior belief of $\tilde\theta$ after that recommendation. 
If types in an open neighborhood of $\tilde\theta$ observe recommendation $p=v_1$, they would have posterior beliefs over an open neighborhood of $\belief(\tilde\theta)$, in accordance with equation \eqref{eq:posterior}. Because pricing indifference curves in $\Delta(V)$ have measure zero, a strictly positive measure of types in $\tilde\Theta$ would strictly prefer to follow that recommendation, and would thus strictly benefit from $\info(\tilde\theta)$. It follows by incentive compatibility that the seller's rents are strictly positive.

Part 2. Assume the stated condition holds. Consider any direct menu that leads to an efficient outcome. There exists $\overline\e>0$ such that for all $0<\e<\overline\e$, type $\type_{\e}$, defined as
\begin{align}
    \type_{\e}\simeq(\e^{n-1},\e^{n-2},\dots,\e^{2},\e,1-\sum_{k=1}^{n-1}\e^k),
\end{align}
belongs to $\Theta$ and prices efficiently after observing  direct signal $\info(\type_{\e})$, where the approximation means being in an $\varepsilon^n$-neighborhood.  

Start with value $v=v_1$. Since $\type_{\e}$ attaches strictly positive probability to $v=v_1$ and prices efficiently, $\info(\type_{\e})$ recommends $p=v_1$ at value $v_1$ with probability 1. For the recommendation to be incentive compatible, this recommendation must be sent with probability $O(\e^{k-1})$ at all values $v_k$, $1<k\leq n$.\footnote{That is, the recommendation probability is bounded by $L\cdot \e^{k-1}$ for some fixed $L$.} Proceed to value $v=v_2$. Because $\type_{\e}$ attaches strictly positive probability to $v=v_2$ and prices efficiently, and $\info(\type_{\e})$ recommends price $p=v_1$ with probability $O(\e)$ at $v=v_2$, it must be that  $\info(\type_{\e})$ recommends price $p=v_2$ with probability $1-O(\e)=O(1)$ at $v=v_2$. For the recommendation to be incentive compatible, it must be sent with probability $O(\e^{k-2})$ at all values $v_k$, $2<k\leq n$. Proceeding analogously for all higher values, we obtain that signal $\info(\type_{\e})$ must take the following form:
\begin{align}
    \begin{array}{cc|cccc}
    \theta_{\e} & \info(\type_{\e}) & p=v_1 & p=v_2 & \dots & p=v_n \\
    \hline \e^{n-1} & v=v_1 & 1 & 0 & \dots  & 0\\
    \e^{n-2} & v=v_2 & O(\e) & O(1) & \dots  & 0\\
    \dots & \dots & \dots & \dots & \ddots  & 0\\
    1-\sum_{k=1}^{n-1}\e^k & v=v_n & O(\e^{n-1}) & O(\e^{n-2}) & \dots  & O(1)
    \end{array}.
\end{align}
When $\e\rightarrow 0$, $\info(\type_{\e})$ converges to a fully informative signal. Since $\e$ can be set arbitrarily small,  incentive compatibility implies that each type $\type\in\types$ earns a maximal possible rent and thus is offered a fully informative signal. 
\end{proof}

\end{document}